\documentclass[review,number,12pt,3p,times]{elsarticle}

\usepackage{amssymb}
\usepackage{amsmath}
\usepackage{eqnarray} 
\usepackage{stackengine}		
\usepackage{scalerel}		
\usepackage{xcolor}			
\usepackage{booktabs}   
\usepackage{siunitx}    
\usepackage{graphicx}
\usepackage{pgfplotstable} 
\usepackage{layouts}
\usepackage[hidelinks]{hyperref}
\usepackage{url}
\usepackage{caption}

\newcommand\openbigstar[1][0.7]{%
  \scalerel*{%
    \stackinset{c}{-.125pt}{c}{}{\scalebox{#1}{\color{white}{$\bigstar$}}}{%
      $\bigstar$}%
  }{\bigstar}
}

\journal{arXiv Preprint}
\begin{document}


\begin{frontmatter}

\title{Broadband Dielectric Analysis of Clays: Impact of Cation Exchange Capacity, Water Content, and Porosity}

\author[a]{Felix Schmidt\fnref{orcidA}}
\author[b]{Norman Wagner\fnref{orcidB}}
\author[c]{Ines Mulder \fnref{orcidC}}
\author[d]{Katja Emmerich\fnref{orcidD}}
\author[e]{Thierry Bore\fnref{orcidE}}
\author[a,f,g]{and Jan Bumberger\fnref{orcidF}}

\fntext[orcidA]{ORCID: 0000-0001-6661-2972}
\fntext[orcidB]{ORCID: 0000-0001-5416-7162}
\fntext[orcidC]{ORCID: 0000-0002-9240-6862}
\fntext[orcidD]{ORCID: 0000-0001-5600-6581}
\fntext[orcidE]{ORCID: 0000-0001-6084-094X}
\fntext[orcidF]{ORCID: 0000-0003-3780-8663}

\affiliation[a]{organization={Department Monitoring and Exploration Technologies, Helmholtz Centre for Environmental Research - UFZ},
                addressline={Permoserstraße 15}, 
                city={Leipzig}, 
                postcode={04315}, 
                country={Germany}}

\affiliation[b]{organization={Institute of Material Research and Testing - MFPA, Bauhaus-University Weimar},
                addressline={Coudraystraße 9}, 
                city={Weimar}, 
                postcode={99421}, 
                country={Germany}}

\affiliation[c]{organization={Department of Geography, Soil Science and Soil Resources, Ruhr-Universität Bochum},
                addressline={Universitätsstraße 150}, 
                city={Bochum}, 
                postcode={44801}, 
                country={Germany}}

\affiliation[d]{organization={Competence Center for Material Moisture - CMM, Applied Mineralogy {/} Clay Science, Karlsruhe Institute of Technology - KIT},
                addressline={Hermann-von-Helmholtz-Platz 1}, 
                city={Eggenstein-Leopoldshafen}, 
                postcode={76344}, 
                country={Germany}}

\affiliation[e]{organization={School of Civil Engineering, University of Queensland},
                addressline={St Lucia}, 
                city={Brisbane}, 
                postcode={4072}, 
                country={Australia}}
								
\affiliation[f]{organization={Research Data Management - RDM, Helmholtz Centre for Environmental Research - UFZ},
                addressline={Permoserstraße 15}, 
                city={Leipzig}, 
                postcode={04315}, 
                country={Germany}}
								
\affiliation[g]{organization={German Centre for Integrative Biodiversity Research (iDiv) Halle-Jena-Leipzig},
                addressline={Puschstraße 4}, 
                city={Leipzig}, 
                postcode={04103}, 
                country={Germany}}

\begin{abstract}
Clay-rich soils and sediments are key components of near-surface systems, influencing water retention, ion exchange, and structural stability. Their complex dielectric behavior under moist conditions arises from electrostatic interactions between charged mineral surfaces and exchangeable cations, forming diffuse double layers that govern transport and retention processes. This study explores the broadband dielectric relaxation of four water-saturated clay minerals (kaolin, illite, and two sodium-activated bentonites) in the 1~MHz -- 5~GHz frequency range using coaxial probe measurements.\\

The dielectric spectra were parameterized using two phenomenological relaxation models -- the Generalized Dielectric Relaxation Model (GDR) and the Combined Permittivity and Conductivity Model (CPCM) -- alongside two theoretical mixture models: the Augmented Broadband Complex Dielectric Mixture Model (ABC-M) and the Complex Refractive Index Model (CRIM). These approaches were evaluated for their ability to link dielectric relaxation behavior to key petrophysical parameters cation exchange capacity (CEC), volumetric water content (VWC), and porosity.\\

The results demonstrate distinct spectral signatures correlating with clay mineralogy, particularly in the low-frequency range. Relaxation parameters, such as relaxation strength and apparent DC conductivity, exhibit strong relationships with CEC, emphasizing the influence of clay-specific surface properties. While expansive clays like bentonites displayed enhanced relaxation due to ion exchange dynamics, deviations in a soda-activated bentonite highlighted the impact of chemical treatments on dielectric behavior.\\

This study provides a systematic framework for linking clay mineral physics with applied electromagnetic methods. The results have significant implications for non-invasive, frequency-domain methods for characterizing soils and sediments, hydrological modeling, geotechnical evaluation, and environmental monitoring.\\
\end{abstract}




\end{frontmatter}





\section{Introduction}

Soils and clay-rich geomaterials are essential components of the Earth's surface system, preserving ecological balance and ensuring human survival. They regulate the planet's water cycle, sequester carbon, and sustain food production for a growing population. However, degradation of these materials and water scarcity increasingly threaten these critical functions, emphasizing the need for advanced monitoring techniques \cite{Cardinale2012, Steffen2015, sdgs2015, Gomiero2016, Gupta2023}.\\

The productivity of clay-dominated subsurface systems relies on soil properties such as nutrient retention, water availability, and structural stability. Among these, cation exchange capacity (CEC) plays a pivotal role in ion transport and surface interactions, directly influencing electrochemical and dielectric behavior under changing environmental conditions. Accurate and scalable assessment methods for these properties remain a significant challenge, particularly for natural, clay-rich porous media where complex interactions dominate \cite{Lin2010, Weil2017, Rodrigues2023}.\\

Understanding the electromagnetic and structural properties of clay-rich materials is crucial for advancements in geophysical characterization, hydrology, and environmental science. Despite substantial research, non-invasive quantification methods for parameters such as CEC, volumetric water content (VWC), and porosity remain underdeveloped. These limitations are particularly evident in soils and sediments containing clays, where specific surface area and electrochemical activity significantly influence dielectric response and ion dynamics \cite{Weil2017, Gupta2023, Vogel2024}.\\

The effective CEC reflects a soil’s capacity to retain and exchange positively charged ions at a given soil pH, which is essential for plant nutrition, water retention, and soil structural stability. Clay minerals, as primary constituents of soils, are – besides soil organic matter - critical in determining the CEC due to their high surface area and potentially high permanent layer charge. Quantifying CEC has broad implications for agricultural productivity, ecosystem health, and climate resilience strategies \cite{Gomiero2016, Guo2016, Rodrigues2023}.\\

Beyond CEC, parameters such as VWC and porosity are equally vital. These properties influence hydrological modeling, soil stability, plant growth, and groundwater recharge. The application of dielectric relaxation spectroscopy offers a unique, non-invasive method to study these interrelated properties (CEC, VWC, and porosity), enhancing our understanding of clay-rich geomaterial behavior under diverse environmental conditions \cite{Lin2010, Cardinale2012, Vogel2024}.\\

Traditional methods for assessing soil and sediment properties, such as chemical extractions and gravimetric measurements, are often time-consuming and unsuitable for large-scale or in-field applications. To overcome these limitations, broadband high frequency (radio to microwave) electromagnetic (HF-EM) methods present a promising alternative. Although primarily applied in laboratory settings in this study, these methods are being explored for their potential in future in-field applications. They leverage the relationship between dielectric permittivity and soil properties, enabling non-invasive, rapid, and scalable analyses. Previous research has shown that dielectric spectra capture the interactions between clay mineral surfaces and aqueous pore solutions, facilitating indirect estimation of parameters like CEC, VWC, and porosity \cite{Weil2017, WagScheu2017}.\\

Despite the progress in dielectric spectroscopy, significant challenges persist. The dielectric relaxation behavior of clay minerals is influenced by a multitude of complex factors, including mineralogical composition, water content, porosity, and ion exchange dynamics. Interpreting broadband dielectric spectra (spanning MHz to GHz) requires robust modeling approaches capable of isolating the contributions of overlapping relaxation processes and connecting them to critical material-specific physico-chemical parameters. In this study, we used various clays commonly found in soils as simplified soil systems to systematically investigate these relationships.\\

The objectives of this study are as follows: 
\begin{enumerate} 
	\item Experimentally determine the broadband dielectric relaxation spectra of water-saturated expansive and non-expansive clays. 
	\item Decompose the observed spectra into distinct relaxation processes using phenomenological and theoretical models. 
	\item Evaluate the potential of these models to estimate critical clay properties, including CEC, VWC, and porosity.
	\item Compare the applicability of the proposed methods for advancing geomaterial characterization, with a focus on potential applications in subsurface exploration, hydrological modeling, and geotechnical analysis.\\
\end{enumerate}

This work specifically examines kaolin, illite, and two sodium-activated bentonites to investigate the relationships between dielectric properties, CEC, VWC, porosity, and clay mineralogy. Additionally, it evaluates the performance of three modeling approaches: the Generalized Dielectric Relaxation Model (GDR), the Combined Permittivity and Conductivity Model (CPCM), and the Augmented Broadband Complex Dielectric Mixture Model (ABC-M). A comparative analysis highlights the strengths, limitations, and suitability of these models for improving geophysical and material-focused characterization methods.\\

In summary, this research aims to bridge the gap between fundamental clay mineral physics and applied electromagnetic techniques. By providing a foundation for non-invasive, frequency-domain dielectric analysis, it supports improved modeling of clay-rich systems in geotechnical engineering, environmental geophysics, and near-surface exploration.


\section{Broadband HF-EM Methods and Models for Soil Dielectric Properties}

\subsection{Overview of HF-EM Methods for Soil Characterization}

HF-EM methods include ground penetration radar (GPR), time domain reflectometry (TDR), various remote sensing methods, borehole logging methods, and various soil moisture sensors. These methods are mostly based on the measurement of dielectric permittivity and a model that links this to soil properties such as soil water content, CEC, porosity, matrix potential, and soil electrical conductivity \cite{WagScheu2017}.\\

One interesting case is the estimation of the CEC using dielectric permittivity. CEC describes the soil's ability to retain nutrients and is therefore of particular importance in agricultural applications. The electromagnetic response caused by the amount of cations that can be exchanged at clay mineral surfaces has been primarily investigated in the low-frequency (LF) range \cite{Revil2013}. However, measurements in the mHz range are often time-consuming and susceptible to electrode polarization effects in the upper LF range. In contrast, in the radio frequency (RF) range, these effects are less relevant, and in some cases, non-contact and spatially resolved measurements using radar or remote sensing methods are feasible. Nevertheless, the precise mechanisms influencing the RF range, particularly for clay-rich soils, are not yet fully understood and require further investigation.\\

Multi-phase porous materials, such as soil or rock containing clay minerals, typically exhibit several distributed relaxation processes in the application-relevant frequency range HF-EM from 1~MHz to approximately 10~GHz \cite[]{Stroud1986, Ishi00, Assifaoui2001, Ishi03, Kell05, gonzalez2020dielectric}. Dispersion and absorption in the frequency range below 1~GHz are dominated by the fine grain fraction ($<$~2~$\mu$m) of the material \cite[]{Arcone2008}, i.e., the clay size fraction and thus clay mineralogy.\\

From a geomechanical point of view, soils containing clay minerals are challenging because of their low strength, high compressibility, and high level of volumetric changes (swelling potential, \cite[]{ural2018importance}). Clay particles range in diameter from 2~$\mu$m downward and constitute the colloidal soil fraction; they are plate- or needle-like and belong mainly to the aluminosilicates. The clay fraction strongly affects soil physico-chemical behavior because of its high surface area per unit mass and physicochemical activity \cite[]{verheye2009land}.\\

For some clay minerals, isomorphous substitution of atoms within a crystal structure creates a negative structural charge that attracts positively charged (exchangeable) cations. Under moist conditions, these ions dissociate from the clay particles. The hydrated colloidal clay forms a structure in which the adsorbed ions are spatially separated from the negatively charged particle surface to the bulk solution, forming an electrostatic double layer \cite[]{schulze1989introduction, verheye2009land, israelachvili2015intermolecular}. The total number of negative charges that can be balanced by exchangeable cations on the clay particle surface is nearly constant under chemically neutral conditions and is independent of the number of cations present in the aqueous pore solution, which is commonly referred to as the potential CEC. A distinction must be made between the CEC of cations on the clay particle surface due to the layer charge of the clay minerals and the deposition on the edges of the particles. Both charges together result in the total number of cations that can be deposited or cation exchange capacity. The deposition due to the edge charge is dependent on the pH value. In swellable clay minerals, such as montmorillonite, the interlayer charge is dominant, and the edge charge accounts for only 10–12\% of the CEC. For non-swellable minerals, such as illite and kaolinite, the CEC only reflects the pH-dependent edge charge value. Therefore, the CEC characterizes an intrinsic property of the soil material that depends only on the pH value \cite[]{astm2010standard}.\\

The dielectric relaxation behavior of porous media contains valuable information about the material due to a strong correlation with the volume fractions of the soil phases as well as contributions by interactions between the pore solution and mineral particles \cite[]{wagner2011, han2012continous, Revil2013, josh2014dielectric, Wagner2014, Josh2015, loewer2017ultra}. Against this background, Garrouch \cite{garrouch1994influence} proposed an approach for the estimation of CEC from dielectric dispersion in well logs.\\

There are only few systematic broadband radio to microwave (kHz to MHz) experimental investigations of saturated clays under defined mechanical conditions available for a characterization of the relaxation behavior. A decomposition of the dielectric relaxation spectra into the underlying relaxation processes is still incomplete and thus the link between the dielectric response and the petrophysical and chemical properties of soils and rocks containing clay remains insufficiently understood \cite{mendieta2021spectral}. An essential reason for this is the necessary combination of different measuring techniques and measuring cells \cite{bore2022experimental}. In addition, individual investigations on specific soils and rocks are focused on specific temperature and frequency ranges for appropriate applications \cite{loewer2017ultra, bore2021analysis, bore2021coupled}. Moreover, there is a lack of protocols for experimental procedures for (i) the preparation of the soft geomaterials, i.e., clay soils, and (ii) installing the prepared material in measurement cells under defined hydraulic and mechanical conditions \cite{lauer2012new, Wagner2013, schwing2016radio, Bore2024}.\\

In general, dielectric investigations can be performed in the frequency domain or in the time domain \cite{Jonscher1996, Kremer2003, kaatze2006, Kaatze2010}. In the low-frequency range $<$ 1 MHz, two- or four-electrode measuring cells are usually used in combination with impedance analyzers \cite{Kule94, Boer06, Asan07, leroy2009mechanistic, wagner2011}. In the high-frequency range $>$ 1 MHz, the coaxial waveguide technique (coaxial sensing elements or open coaxial sensing elements) is used in combination with network analyzers \cite{Ishi03, Kaatze2010, lauer2012new, Wagner2013, schwing2016radio, bumberger2018, bore2021coupled, Schmidt2021}. The relaxation behavior of partially saturated and saturated soils usually shows a significant deviation from the simple Debye behavior \cite{Deby41} and a wide distribution of relaxation processes \cite{Hoek74, chelidze1999electrical, Ishi03, Asan07, zhang2020frequency, gonzalez2020dielectric, bore2021analysis, bobrov2021dielectric}. Therefore, the parameterization is based on corresponding dielectric relaxation models (Cole-Cole \cite{Cole41}, Cole-Davidson \cite{Davi51}, Havriliak-Negami \cite{Havr96}, Jonscher \cite{Jonscher1996}, KWW \cite{Kohl47}, Williams and Watts \cite{Will70}) taking into account the expected relaxation processes in the pore solution as well as processes due to the interaction between the pore solution and the solid \cite{Asan07}.\\

\subsection{Modeling Broadband HF-EM Material Properties}

\begin{figure}[ht!]
	\centering
	\includegraphics[width=\textwidth]{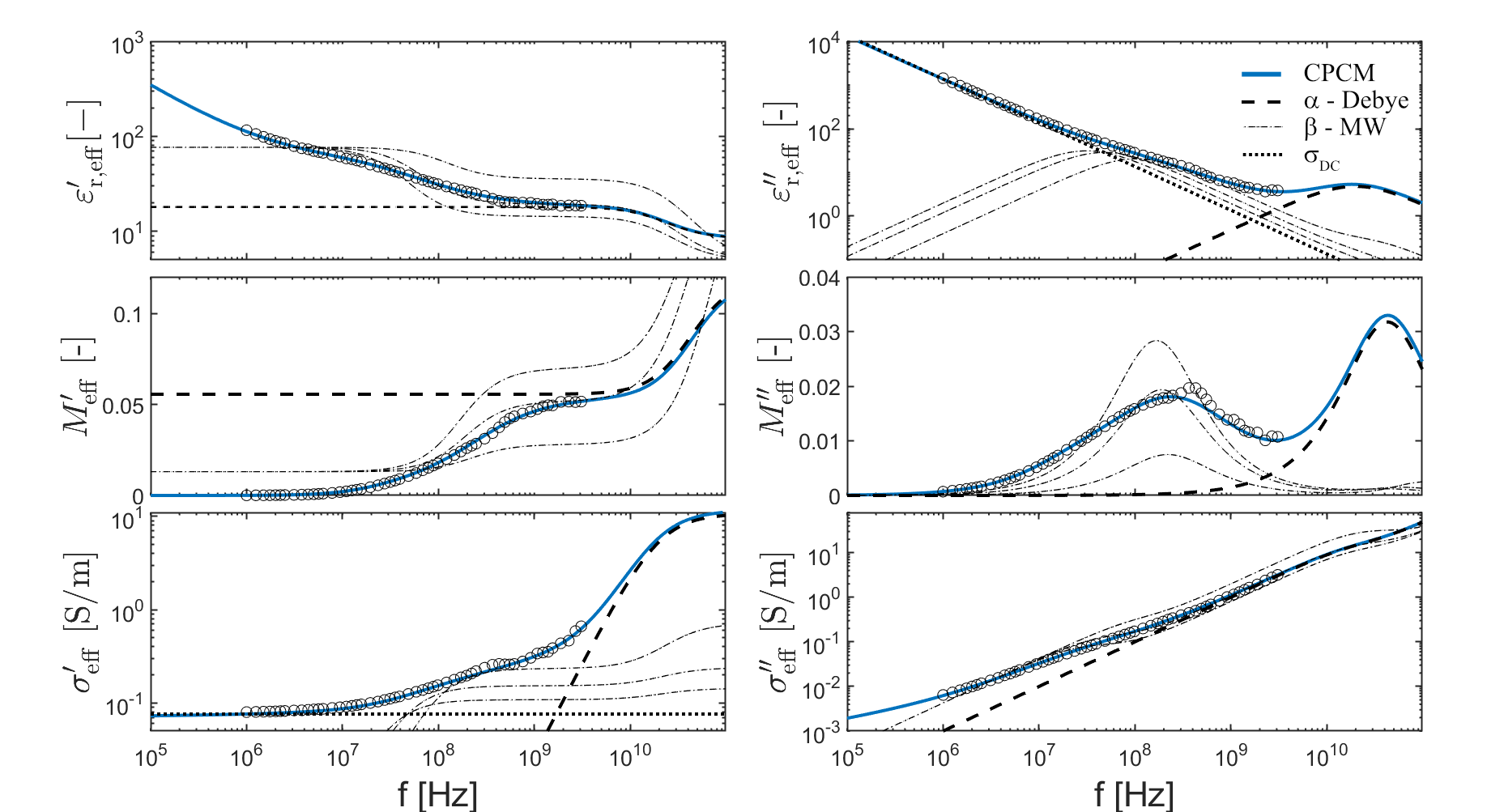}
	\caption{From top to bottom: Real and imaginary part of the complex effective permittivity $\varepsilon^\star_{\mathrm{r,eff}}$ of a nearly water saturated loess soil (c.f. \cite{loewer2017ultra}, volumetric water content $\theta=0.36~\mathrm{m^3/m^3}$, porosity $\phi=0.37$, water saturation $S_{\mathrm{W}}=0.98$), complex electrical modulus $M^\star$ and complex effective conductivity $\sigma^\star_{\mathrm{eff}}$ as a function of frequency. The line represents a theoretical prediction of the measured spectra by means of the Complex Permittivity and Conductivity model (CPCM) according to \cite{loewer2017ultra}. The dashed line is computed with a Debye-type relaxation function (process P1), dashed-dotted line is a single Maxwell-Wagner effect (c.f. \cite{bore2021analysis}) calculated based on the approach by Bona et al. 1998 \cite[]{bona1998characterization}, and dotted line indicates the direct current conductivity contribution to the spectra.}
	\label{fig:SpectraIntro}
\end{figure}

HF-EM properties of non-ferromagnetic clay soils are characterized by a complex, temperature $T$, pressure $p$, and frequency $f$ dependent, effective relative permittivity $\varepsilon^\star_{\mathrm{r,eff}}(T, p, \omega)$, electrical modulus $M^\star_{\mathrm{eff}}(T, p, \omega)=\varepsilon^\star_{\mathrm{r,eff}}(T, p, \omega)^{-1}$, or effective electrical conductivity 
$\sigma^\star_{\mathrm{eff}} = j\omega\varepsilon_{0}\varepsilon^\star_{\mathrm{r,eff}}$, with angular frequency $\omega = 2\pi f$ \cite[]{loewer2017ultra}. These properties are essential for understanding the dielectric response of soils under varying environmental and mechanical conditions. Typical spectra are presented in Figure \ref{fig:SpectraIntro} (\cite{loewer2017ultra}).\\

In general, soils and rocks containing clay minerals exhibit several distributed relaxation processes. These processes are key to interpreting dielectric spectra and understanding the complex interaction between soil phases and electromagnetic waves.\\

\subsubsection{Phenomenological Relaxation Models}

In order to model broadband dielectric spectra of soils and rocks, a generalized dielectric relaxation model (GDR) based on Cole-Cole terms can be applied \cite[]{wagner2011, Wagner2013, loewer2017ultra}:

\begin{equation}
    \varepsilon^\star_{\mathrm{r,cc}} - \varepsilon_{\infty} =  \sum_{i=1}^{k} \frac{\Delta\varepsilon_i}{1+(j\omega\tau_i)^{b_i}}- j\frac{\sigma'_{\mathrm{DC}}}{\omega\varepsilon_0}
\label{eq:cole}
\end{equation}

Here, $\varepsilon_0$ is the dielectric permittivity of vacuum, $\varepsilon_{\infty}$ is the relative constant permittivity at the high frequency limit, $\Delta\varepsilon_{i}$ is the relaxation strength, $\tau_i$ is the relaxation time, $0 \leq b_i \leq 1$ is the stretching exponent of the $i$-th relaxation process, and $\sigma'_{\mathrm{DC}}$ is the apparent direct current conductivity.\\

There is strong evidence that HF-EM properties above 1~GHz are dominated by water relaxation processes, which can be modeled using a single Debye term ($b_{1,\mathrm{water}} = 1$, $\alpha$-process) \cite{Robinson2003, wagner2011, Wagner2013, Wagner2014, loewer2017ultra}. Dispersion and absorption at frequencies below approximately 1~GHz are typically modeled with two additional Cole-Cole terms ($\alpha'$, $\beta$) to account for interfacial relaxation processes \cite{wagner2011, Wagner2013}. This approach results in $k=3$ for equation (\ref{eq:cole}).\\

When using the effective relative permittivity $\varepsilon^\star_{\mathrm{r,eff}}(\omega)$ to characterize relaxation processes, it becomes evident that DC-conductivity dominates the relaxation behavior in the range of $f \leq 1~\mathrm{MHz}$. For low-frequency (LF, $f \leq 1~\mathrm{MHz}$) modeling, it is often advantageous to use the effective electrical conductivity $\sigma^\star_{\mathrm{eff}}$ instead. This approach avoids the high values of relaxation strength associated with permittivity-based models in the LF range. Moreover, polarization parameters such as the chargeability of specific processes can be related to soil properties like CEC using the Dynamic Stern Layer model \cite{Revil2017complex}. However, permittivity-based models are more suitable for analyzing HF relaxation processes and can be effectively linked to theoretical mixture models \cite{wagner2011, Wagner2013}.\\

To combine these advantages, the Combined Permittivity and Conductivity Model (CPCM) was proposed by Loewer et al. \cite{loewer2017ultra}. This model integrates both permittivity- and conductivity-based approaches to cover the entire frequency spectrum. It includes a Debye term for water relaxation at high frequencies, a permittivity-based Cole-Cole term for intermediate frequencies, and a conductivity-based Cole-Cole term for low frequencies:

\begin{equation}
\begin{aligned}
\varepsilon^{*}_{\mathrm{r,eff}} = & \varepsilon_{\infty} + \frac{\Delta\varepsilon_1}{1+(j\omega\tau_1)^{c_1}} + \frac{\Delta\varepsilon_2}{1+(j\omega\tau_2)^{c_2}} \\
& -j\frac{\sigma'_{\mathrm{dc}}}{\omega\varepsilon_{0}} \left[ 1 + \frac{M'}{1-M'} \left(1 - \frac{1}{1+(j\omega\tau_3)^{c_3}}\right) \right]
\end{aligned}
\label{eq:cpcm_summe}
\end{equation}

Here, $0 \leq c_i \leq 1$ are stretching exponents, and $M'$ represents the apparent chargeability. Chargeability is linked to the DC-conductivity $\sigma_{\mathrm{DC}}$ and instantaneous conductivity $\sigma_{\infty}$ through the relationship $\sigma_{\mathrm{DC}} = \sigma_{\infty} (1-M)$ \cite{Revil2017complex}.\\

In this study, apparent HF-chargeability $M'$ is estimated using the relaxation mode as described by Weigand et al. \cite{weigand2016debye}. This hybrid modeling framework allows for the characterization of both HF and LF processes, enabling more accurate interpretation of dielectric spectra across wide frequency ranges.\\

\begin{figure}[htb]
	\centering
	\includegraphics[width=0.5\textwidth]{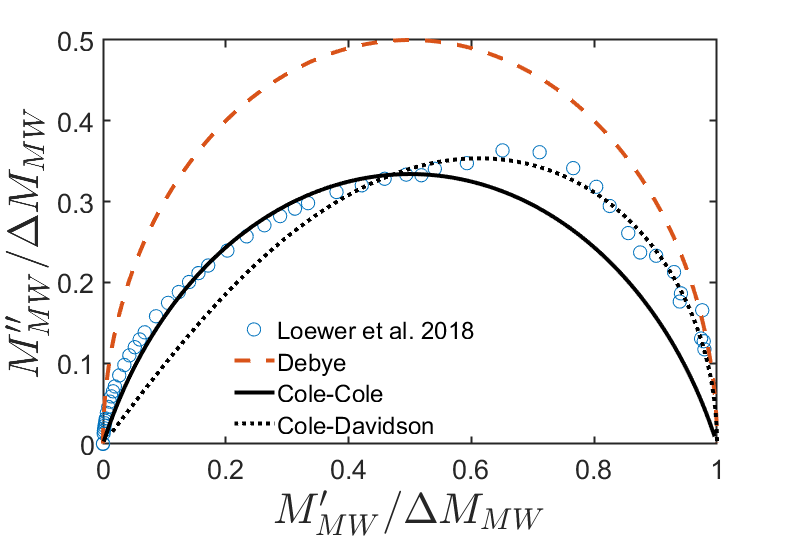}
	\caption{Representation of the complex modulus for Maxwell-Wagner polarization normalized with the modulus strength between 1~MHz and 1~GHz, $\Delta M_{\mathrm{MW}} = \varepsilon'^{-1}_{\mathrm{r, eff}}(f = 1~\mathrm{GHz})$, in the Gaussian plane.}
	\label{fig:Empirical}
\end{figure}

Figure \ref{fig:Empirical} illustrates the Maxwell-Wagner polarization in the complex modulus representation. A pure Maxwell-Wagner process leads to a Debye-type relaxation, while distributed processes produce more generalized behaviors. These can be modeled using the Havriliak-Negami-type modulus representation \cite{Havr96}:

\begin{equation}\label{eq:HN}
    \frac{M^\star_i}{\Delta M_i} = \frac{(j\omega\tau_i)^{\alpha_i \cdot \beta_i}}{\left(1 + (j\omega\tau_i)^{\alpha_i}\right)^{\beta_i}}
\end{equation}

Here, $\Delta M_i$ represents the modulus strength of the $i$-th process, $\tau_i$ the relaxation time, and $0 \leq \alpha_i, \beta_i \leq 1$ the stretching exponents. For $\alpha_i = \beta_i = 1$, equation (\ref{eq:HN}) reduces to a Debye function. For $\alpha_i = 1$, it simplifies to a Cole-Davidson function, and for $\beta_i = 1$, it becomes a Cole-Cole function.\\

\subsubsection{Theoretical Mixture Models}

Theoretical mixture models provide a robust framework for describing the effective dielectric properties of soils, considering their multiphase composition. A widely used approach is the Complex Refractive Index Model (CRIM), which relates the effective permittivity to the individual contributions of soil phases \cite[]{Wang80, Arcone2008, Scho2015}:

\begin{equation}
    \varepsilon^{\star~0.5}_\mathrm{r,mix} =\theta \varepsilon^{\star~0.5}_{\mathrm{r,W}} + (1-\phi)\varepsilon_{\mathrm{G}}^{0.5} + (\phi-\theta)
\end{equation}

Here, $\theta$ represents the volumetric water content, $\phi$ the porosity, and $\varepsilon_{\mathrm{G}}$ the permittivity of the solid particles.\\

The electromagnetic properties of the solid particles are, strictly speaking, a second-order tensor (dyadic) with nine independent material parameters \cite[]{Scho2015}. However, for practical applications, the relative effective permittivity of the solid matrix material $\varepsilon_{\mathrm{G}}$ can be estimated based on the mineralogical composition of the solid phases, assuming quasi-isotropy at the sample scale \cite[]{Robi04a, Robinson2004, Robinson2003b}. A lack of systematic broadband frequency- and temperature-dependent high-resolution electromagnetic investigations under defined stress-strain conditions on mono- and poly-crystalline mineral phases remains a significant limitation \cite{Jones2000a, Boivin2022a, Boivin2022b}. Addressing this issue is crucial for advancing the accuracy of theoretical mixture models.\\

To estimate $\varepsilon_{\mathrm{G}}$, several empirical relationships have been proposed that link the relative effective permittivity of a material to its grain density $\rho_G$ (in $\mathrm{g/cm^3}$). Notable examples include:

\begin{equation}\label{eq:olhoeft}
\varepsilon_{\mathrm{G}}=A^{\rho_G}
\end{equation}

as suggested by Olhoeft \cite{Olho74}, with $A=1.93\pm0.17$, or by Campbell \cite{Campbell2002}, with $A=1.96$. Another widely used empirical equation was introduced by Dobson \cite{Dobson1985}:

\begin{equation}\label{eq:Dobson}
\varepsilon_{\mathrm{G}}=(1.01+0.44\cdot\rho_G)^2-0.062
\end{equation}

In this study, $\varepsilon_{\mathrm{G}}$ was determined as the geometrical mean of equations (\ref{eq:olhoeft}) and (\ref{eq:Dobson}).\\

To compute the complex permittivity of the aqueous pore solution $\varepsilon^\star_{\mathrm{r,W}}$, a modified Debye model was employed:

\begin{equation}\label{eq:Debye}
\varepsilon^\star_{\mathrm{r,W}} -  \varepsilon_{\infty}=   
\frac{\varepsilon_{\mathrm{s,W}}-\varepsilon_{\infty}}{1+j\omega\tau_{\mathrm{W}}} - j\frac{\sigma_{\mathrm{W}}}{\omega\varepsilon_0}
\end{equation}

Here, $\varepsilon_{\infty}$ represents the high-frequency limit, $\varepsilon_{\mathrm{s,W}}$ the quasi-static relative permittivity, and $\tau_{\mathrm{W}}$ the relaxation time. Debye parameters were calculated using equations from \cite{kaatze2007} and \cite{Buchner1999}. The direct current (DC) conductivity of the pore water, $\sigma_{\mathrm{W}}$, was treated as an unknown parameter determined via inverse modeling (see Section~\ref{sec:Inversion}).\\

In addition to the CRIM model, the Augmented Broadband Complex dielectric mixture Model (ABC-M) was applied, as suggested in \cite[]{bore2018}:

\begin{equation}
    \varepsilon_{\mathrm{r,eff}}^\star(\omega) =\frac{\Delta\varepsilon_{\mathrm{cc}}}{1+(j\omega\tau_{\mathrm{cc}})^{b_{\mathrm{cc}}}} +  \varepsilon_{\mathrm{r, mix}}^\star
\end{equation}

It combines a relaxation term with a mixture equation:

\begin{equation}\label{eq:RevMix}
    \varepsilon_{\mathrm{r, mix}}^\star= \phi^m[S^n_{\mathrm{W}} \cdot \varepsilon_{\mathrm{r,W}}^\star +
    (1-S^n_\mathrm{W}) + (\phi^{-m}-1)\varepsilon_{\mathrm{G}}]
\end{equation}

Here, $S_{\mathrm{W}} = \theta / \phi$ denotes water saturation, while $n$ and $m$ are the saturation and cementation exponents, respectively. This combination of relaxation and mixture terms makes ABC-M a versatile tool for analyzing soil dielectric properties under diverse conditions, particularly for clay-rich soils where complex interactions dominate. Here, $n$ and $m$ can be approximated as $n \approx m = m_0 + \mathrm{CEC} \cdot m_1$, with $m_0 = 1.8$ and $m_1 = 4.3 \cdot 10^{-5}~\mathrm{kg/C}$. This simplifies equation (\ref{eq:RevMix}) as follows:

\begin{eqnarray}
  \sigma'_{\mathrm{eff}} &\approx & \theta^m\sigma'_{\mathrm{w}}+\theta^{m-1}\sigma'_{\mathrm{s}}
  \label{eq:revil_formula_sigma}\\
  \varepsilon^\star_{\mathrm{r,eff}} &\approx &
  \theta^m\varepsilon^\star_{\mathrm{w}}+(1-\theta^m)\varepsilon_{\mathrm{G}}
  \label{eq:revil_formula_epsilon}
\end{eqnarray}

In these equations, $\sigma'_{\mathrm{eff}}$ represents the effective conductivity, $\sigma'_{\mathrm{w}}$ the real electrical conductivity of the pore fluid, $\sigma'_{\mathrm{s}} = \sigma_{s, 1.3~\mathrm{GHz}} = (0.44 \pm 0.1)~\mathrm{S/m}$, according \cite{Revil2013}, the real surface conductivity contribution, and $\varepsilon^\star_{\mathrm{w}}$ the complex permittivity of the pore solution.\\

\subsubsection{Empirical and Semi-Empirical Approaches}

Early studies extensively explored the empirical relationship between soil water content and high-frequency ($\approx$1~GHz) relative permittivity. A foundational empirical equation relating volumetric water content $\theta$ to apparent relative permittivity $\varepsilon_{\mathrm{A}}$ was proposed by Topp \cite{Topp1980} based on Time Domain Reflectometry (TDR) measurements with coaxial transmission lines:

\begin{equation}\label{eq:topp}
\varepsilon_{\mathrm{A}} = 3.03 + 9.3 \cdot \theta + 146.0 \cdot \theta^2 - 76.7 \cdot \theta^3
\end{equation}

This equation has been widely used for characterizing soil water content due to its simplicity and effectiveness in many soil types.\\

Wensink \cite{wensink1993dielectric} systematically investigated the frequency dependence of relative effective permittivity $\varepsilon_{\mathrm{r,eff,W}} = \varepsilon'_{\mathrm{r,eff}}$ and effective electrical conductivity $\sigma_{\mathrm{eff,W}} = j\omega\varepsilon_0\varepsilon''_{\mathrm{r,eff}}$ over a range from 1~MHz to 3~GHz. Empirical equations for specific frequencies, including 5~MHz, 50~MHz, and 1~GHz, were derived. For this study, the equation for 1~GHz is particularly relevant:

\begin{equation}\label{eq:wensink}
\varepsilon_{\mathrm{r,eff,WS}}(f = 1~\mathrm{GHz}) = 3.2 + 41.4 \cdot \theta + 16.0 \cdot \theta^2
\end{equation}

This empirical approach effectively correlates volumetric water content with relative permittivity at high frequencies.\\

Josh et al. \cite{Josh2015} introduced a method to determine the dispersion of the real part of the permittivity by calculating the difference between its values at 1~GHz and 1~MHz:

\begin{equation}\label{eq:Josh}
\Delta\varepsilon_{\mathrm{1GHz}} = \varepsilon'_{\mathrm{r,eff}}(f = 1~\mathrm{MHz}) - \varepsilon'_{\mathrm{r,eff}}(f = 1~\mathrm{GHz})
\end{equation}

This approach enables the characterization of superimposed relaxation processes with relaxation frequencies below the bulk water relaxation frequency. However, it does not establish a direct relationship between soil physical properties and specific relaxation mechanisms. Instead, it identifies correlations between dispersion behavior and clay mineralogy.\\

In addition to dispersion analysis, direct current (DC) conductivity contributions can be determined from the spectral plateau of the measured real part of the complex conductivity (refer to Figure \ref{fig:SpectraIntro}). The depression of the imaginary part in the modulus representation within the complex plane provides further insight into distributed relaxation processes, such as Maxwell-Wagner polarization, physical bound water features, and counter-ion effects.\\


\section{Materials and Methods}

To cover a CEC range of \num{4.4} - \num{110} \si{cmol(+).kg^{-1}}, four different clay samples were selected, each containing typical primary clay minerals (see Table \ref{tab:sample_data_xrd}). The mineral proportions were determined by X-ray diffraction (XRD), confirming the dominant clay minerals in each sample. XRD was conducted using a Bruker D8 Advance A25 diffractometer, equipped with a LynxEye XE detector (Bruker AXS GmbH, Karlsruhe, Germany). Cu-K$\alpha$ radiation was used to record the diffraction patterns, and Rietveld refinement was carried out using the open-source program Profex \cite{doeblin2015profex}.\\

The two bentonite samples were activated with sodium carbonate, converting calcium bentonite into sodium bentonite. This process is commonly used in industrial applications to prevent rapid sedimentation of clay materials in water suspensions. According to the manufacturer, bentonite 1 received more sodium carbonate than bentonite 2, resulting in overactivation. Consequently, bentonite 1 contains an oversupply of ions in suspension that are not adsorbed on the clay mineral surfaces.

\begin{table*}[ht]
  \begin{center}
	\caption{
		The X-ray diffraction (XRD) data of the four clays
 	}
    \label{tab:sample_data_xrd}
    \begin{scriptsize}
    \pgfplotstabletypeset[
	columns={
		clayName,
		fraction,
		mineralName
	},
      col sep=comma, 
      display columns/0/.style={
		column name=clay,
		column type={S},
		string type,
		column type=l
	},
	display columns/1/.style={
		column name=\%,
		column type={S},
		string type,
		column type=r
	},
	display columns/2/.style={
		column name=minerals,
		column type={S},
		string type,
		column type=l
	},
	every head row/.style={
		before row={\toprule}, 
	},
	every last row/.style={after row=\bottomrule}, 
	every row no 0/.style={before row=\midrule},
	every row no 2/.style={after row=\midrule},
	every row no 7/.style={after row=\midrule},
	every row no 13/.style={after row=\midrule},
    ]{
clayName,fraction,mineralName
kaolin,82,kaolinite
,9,dioctahedral mica
,9,quartz; traces of accompanying minerals
illite,68,dioctahedral illite 1M
,11,kaolinite
,6,trioctahedral mica
,5,dioctahedral mica
,10,calcite; feldspar (microcline); quartz; hematite
bentonite1,67,Na-saturated dioctahedral smectite (montmorillonite)
,13,dioctahedral mica
,7,quartz
,6,calcite
,5,dolomite
,2,trioctahedral mica;  feldspar (microcline)
bentonite2,92,Na-saturated dioctahedral smectite (montmorillonite)
,8,aragonite; calcite; feldspar (plagioclase); quartz
} 
  \end{scriptsize}
  \end{center}
\end{table*}


\subsection{Material Preparation}

\begin{figure}[ht!]
	\centering
	\includegraphics[width=0.53\linewidth]{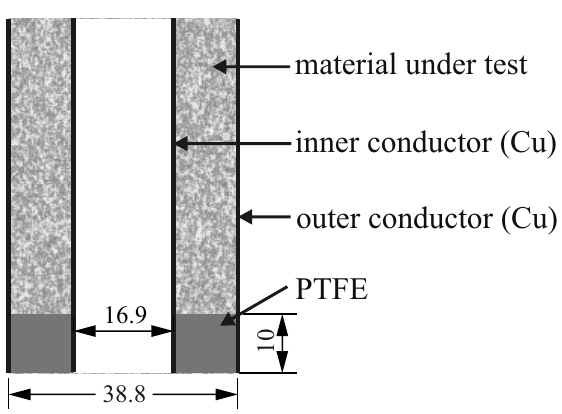}
	\caption{Cross-section of the complete coaxial cell sample holder, featuring PTFE rings at the lower end. The upper end is left unsealed. Dimensions are given in millimeters.}
	\label{fig:coax_cell_cut_vertical}
\end{figure}

The dry samples of illite and bentonites were mixed with deionized water until saturation, exceeding the liquid limit of the clay. This ensured that air-filled macro-pores were avoided. The sample holder, a 60~mm-long coaxial transmission line, was sealed at the lower end with a polytetrafluoroethylene (PTFE) ring, as shown in Figure \ref{fig:coax_cell_cut_vertical}. Detailed construction and dimensions of the sample holder are described in \cite{lauer2012new}.\\

\begin{figure}
	\centering
    \includegraphics[height=0.42\linewidth,width=0.42\linewidth]{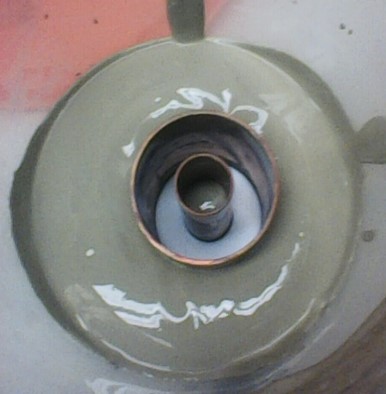}
	\includegraphics[height=0.42\linewidth,width=0.305\linewidth]{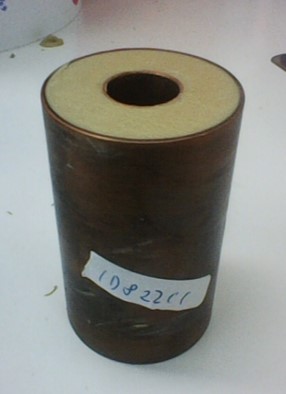}
	\caption{Left: Empty coaxial cell with a PTFE ring inserted into a clay-water suspension. Right: Prepared bentonite sample in the coaxial cell, prior to insertion into the measurement adapter.}
	\label{fig:sample_preparation}
\end{figure}

The sample holder was placed on top of the suspension and allowed to sink freely into the material due to gravity (Figure \ref{fig:sample_preparation}, left). The samples were then left to dry and shrink at room temperature for several days. To prevent the formation of cracks and voids, the suspension container was placed on a vibration table and shaken every two hours for 15 minutes.\\

When the material began forming cracks that did not close upon shaking, the sample holder was removed, cleaned, and the surface of the material was leveled (Figure \ref{fig:sample_preparation}, right). This preparation resulted in fully saturated, homogeneous clay/water mixtures that remained stable when the holder was inverted.\\

Samples of illite 1 and illite 2 were derived from the same raw material. However, illite 1 was dried for a longer duration, resulting in lower water content and reduced porosity.\\

Vertical inhomogeneities in water content can occur when the sample is stored with the PTFE ring pointing downward. To mitigate this, the filled sample holder was sealed airtight and stored with the PTFE ring facing upwards. The sealing was later removed, and the sample was inserted into the vector network analyzer (VNA) adapter unit (Figure \ref{fig:coax_cell_photo}). The right adapter was mounted on a sliding rail, allowing parallel movement relative to the left adapter. This setup prevented lateral forces on the sample and ensured proper alignment of the inner and outer conductors.\\

The influence of the PTFE rings was subsequently removed using high-frequency (HF) de-embedding techniques and the dielectric properties of PTFE. In the frequency range used, PTFE is approximated as $\varepsilon_{\mathrm{r}}' = 2.1$ and $\varepsilon_{\mathrm{r}}'' = 0.003$. The homogeneity of the samples was verified by comparing the scattering parameter reflection factors from the left ($S_{11}$) and right ($S_{22}$) sides, which should match for homogeneous samples.\\

\begin{scriptsize}
\begin{table*}[ht]
  \begin{center}	
\caption{Physical parameters of the 15 investigated clay samples. The water saturation is determined to be $S = 1 (\pm 0.04)$. 
    $\sigma_{\mathrm{w}}$: direct current conductivity of the pore water, inverted with the Complex Refractive Index Model (CRIM); 
    $\sigma_{\mathrm{DC,fit}}$: direct current conductivity fitted to low-frequency data (Eq. \ref{eq:sigma_dc_fit}); 
    $\phi$: porosity; 
    $\theta$: volumetric water content; 
    $w$: gravimetric water content; 
    ID: sample identifier.
}
    \label{tab:sample_data}
    \begin{scriptsize}
    \pgfplotstabletypeset[
	columns={
	clay material,
	dc conductivity,
	sigma_dc_fit,
	porosity,
	volumetric,
	gravimetric,
	ID},
  col sep=comma, 
  display columns/0/.style={
		column name=clay name,
		column type={S},
		string type,
		column type=l},
	display columns/1/.style={
		column name=$\sigma_{\mathrm{w}}$,
		numeric type,
		column type=l,
		fixed zerofill,
		precision=2},
	display columns/2/.style={
		column name=$\sigma_{\mathrm{DC,fit}}$,
		numeric type,
		column type=l,
		fixed zerofill,
		precision=2},
	display columns/3/.style={
		column name=$\phi$,
		numeric type,
		column type=l,
		fixed, precision=2},
	display columns/4/.style={
		column name=$\theta$,
		numeric type,
		column type=l,
		fixed,
		precision=2},
	display columns/5/.style={
		column name=$w$,
		numeric type, 
		dec sep align,
		fixed zerofill,
		precision=2},
	display columns/6/.style={
		column name=ID,
		column type={S},
		string type,
		column type=l},
	every head row/.style={
		before row={\toprule}, 
		after row={
				& $\frac{\mathrm{S}}{\mathrm{m}} $
				& $\frac{\mathrm{S}}{\mathrm{m}} $
				& -
				& $\frac{\mathrm{m}^3}{\mathrm{m}^3}$ 
				& 
				& $\frac{\mathrm{kg}}{\mathrm{kg}}$
				& 
				\\ 
				\midrule 
		} 
	},
	every last row/.style={after row=\bottomrule}, 
    ]{clay material,dc conductivity,sigma_dc_fit,porosity,volumetric,gravimetric,ID
kaolin,0.373,0.075027042,0.49,0.46,0.34,ID602
,0.245,0.062700767,0.531,0.555,0.455,ID644
,0.241,0.062914822,0.535,0.55,0.455,ID645
illite 1,0.491,0.117005411,0.52,0.52,0.41,ID626
illite 2,0.194,0.101187802,0.75,0.73,1.09,ID630
,0.188,0.102333456,0.75,0.72,1.09,ID631
bentonite 1,0.756,0.672584073,0.94,0.95,6.18,ID808
,0.672,0.609611688,0.959,0.968,9,ID822
,0.681,0.622273737,0.9695,0.9895,10.333,ID823
,0.696,0.646217533,0.96,0.97,9,ID824
,0.679,0.645480052,0.966,0.99,11.333,ID825
bentonite 2,0.356,0.30668901,0.96,0.96,8.38,ID809
,0.289,0.254784099,0.97,0.97,12.65,ID834
,0.321,0.270249062,0.97,0.96,11.99,ID840
,0.416,0.340429658,0.95,0.94,8.33,ID844
} 
  \end{scriptsize}
  \end{center}
\end{table*}
\end{scriptsize}

The kaolin samples were prepared differently. The clay powder was dried in an oven at \SI{60}{\celsius} and then sprayed with demineralized water until a gravimetric water content of $w = 0.2$ was achieved. This mixture was compacted in a 4-inch mold according to the ASTM-D698-7e1 \cite{astm698} method, ensuring reproducible porosity and density of the water-saturated sample. The material was then cut from the mold with a hydraulic press and placed in the sample holder, as described in \cite{bore2016}.\\

Unlike other samples, the use of a PTFE ring was unnecessary for kaolin. To achieve full saturation, the cut-out sample with the coaxial cell was placed on a ceramic plate in demineralized water for 24 hours. During this period, the sample absorbed water, resulting in slight swelling and increased porosity. Excess material that leaked from the cell was removed before inserting the sample into the VNA fixture.

\begin{table*}[ht]
  \begin{center}
	\caption{Physical parameters and mean values per clay type. The water saturation is determined to be $S = 1 (\pm 0.04)$.
    BET: specific surface area based on Brunauer–Emmett–Teller (BET) theory; 
    $\rho_{\mathrm{G}}$: particle density measured with a helium pycnometer; 
    $\epsilon_{\mathrm{G}}$: permittivity of the solid particles (Eq. (\ref{eq:olhoeft}) and (\ref{eq:Dobson})); 
    $\sigma_{\mathrm{DC,w}}$: direct current conductivity of extracted porewater from two clay-water ratios; 
    $\sigma_{\mathrm{DC,fit}}$: fitted direct current conductivity of the measured samples (Eq. \ref{eq:sigma_dc_fit}); 
    $\sigma_{\mathrm{w}}$: direct current conductivity of the pore water, inverted with the Complex Refractive Index Model (CRIM); 
    $\phi$: porosity; 
    $\theta$: volumetric water content; 
    $w$: gravimetric water content.
}
    \label{tab:sample_data_mean}
    \begin{scriptsize}
    \pgfplotstabletypeset[
	columns={
	clay material,
	CEC,
	BET,
	particle-density-ines-octanol,
	epsilonG,
	DC-cond-Ines1-5,
	DC-cond-Ines1-20,
	DC-cond-RF,
	dc conductivity,
	porosity,
	volumetric,
	gravimetric
	},
      col sep=comma, 
      display columns/0/.style={
		column name=clay name,
		column type={S},
		string type,
		column type=l},
	display columns/3/.style={
		column name=$\rho_\mathrm{G}$,
		column type={S},
		string type,
		column type=l},
	display columns/4/.style={
		column name=$\epsilon_{\mathrm{G}}$,
		numeric type,
		column type=l,
		fixed zerofill,
		precision=2},
	display columns/5/.style={
		column name=$\sigma_{\mathrm{DC}}$(1:5),
		numeric type,
		column type=l},
	display columns/6/.style={
		column name=$\sigma_{\mathrm{DC}}$(1:20),
		numeric type,
		column type=l},
	display columns/7/.style={
		column name=$\bar{\sigma}_{\mathrm{DC,fit}}$,
		numeric type,
		column type=l},
	display columns/8/.style={
		column name=$\bar{\sigma}_{\mathrm{w}}$,
		numeric type,
		column type=l,
		fixed zerofill,
		precision=2},
	display columns/9/.style={
		column name=$\bar{\phi}$,
		numeric type,
		column type=l,
		fixed, precision=2},
	display columns/10/.style={
		column name=$\bar{\theta}$,
		numeric type,
		column type=l,
		fixed,
		precision=2},
	display columns/11/.style={
		column name=$\bar{w}$,
		numeric type, 
		dec sep align,
		fixed zerofill,
		precision=2},
	every head row/.style={
		before row={\toprule}, 
		after row={
				& $\frac{\mathrm{cmol(+)}}{\mathrm{kg}}$
				& $\frac{\mathrm{m}^2}{\mathrm{g}}$
				& $\frac{\mathrm{g}}{\mathrm{cm}^3}$
				& - 
				& $\frac{\mathrm{S}}{\mathrm{m}} $ 
				& $\frac{\mathrm{S}}{\mathrm{m}} $
				& $\frac{\mathrm{S}}{\mathrm{m}} $ 
				& $\frac{\mathrm{S}}{\mathrm{m}} $
				& -
				& $\frac{\mathrm{m}^3}{\mathrm{m}^3}$ 
				& 
				& $\frac{\mathrm{kg}}{\mathrm{kg}}$
				\\ 
				\midrule 
		} 
	},
	every last row/.style={after row=\bottomrule}, 
    ]{clay material,CEC,BET,particle-density-ines-octanol,epsilonG,DC-cond-Ines1-5,DC-cond-Ines1-20,DC-cond-RF,dc conductivity,porosity,volumetric,gravimetric
kaolin,4.4,16,2.63,5.4582,0.00789,0.00535,0.06688,0.286333333,0.518666667,0.521666667,0.416666667
illite 1,18,105,2.67,5.5905,0.01918,0.0125,0.117,0.491,0.52,0.52,0.41
illite 2,,,,,,,0.10176,0.191,0.75,0.725,1.09
bentonite 1,74,51,2.68,5.6241,0.323,0.241,0.63923,0.6968,0.9576,0.9706,9.1692
bentonite 2,110,70,2.65,5.5905,0.41,0.147,0.29303,0.3455,0.965,0.9575,10.3375
}
  \end{scriptsize}
  \end{center}
\end{table*}

\subsection{Physical and Chemical Parameters}

Physical parameters collected in Table \ref{tab:sample_data} and \ref{tab:sample_data_mean} were determined using standard methods. Gravimetric water content was calculated by weighing small sample portions before and after oven drying. Samples were dried at \SI{105}{\celsius} for \SI{24}{h} and re-weighed. Sample ID602 was dried at \SI{60}{\celsius} until its mass stabilized, which likely resulted in slightly lower gravimetric water content than the actual value.\\

Porosity was calculated based on gravimetric measurements, the measurement cell volume, and the particle density provided by the clay supplier. Volumetric water content was derived using the density of water at \SI{21}{\celsius}. Water saturation was calculated as the ratio of water volume to pore volume. With the described preparation, saturation was close to 1, with an estimated error of $\pm 0.04$.\\

Key material properties were further characterized to facilitate comparison with other studies. CEC was measured by the Cu-trien method according to Meier and Kahr \cite{Meier1999} with changes according to Steudel et al. (2009) \cite{Steudel2009}.\\

The specific surface area (SSA) was determined using nitrogen adsorption based on the Brunauer-Emmett-Teller (BET) theory. For non-swellable clay minerals a correlation of SSA with CEC was observed. In swellable clays, such as bentonites, the SSA primarily indicates platelet size of the smectite \cite{delavernhe2015}. Particle density, measured with a helium pycnometer, reflected the density of the solid components.\\

To compare DC pore water conductivities among the clays, demineralized water was added well above saturation. Clay-to-water mixing ratios of 1:5 and 1:20 were prepared to observe the effect of soda activation. Centrifugation was used to extract pore water, and its conductivity was measured. Higher CEC correlated with higher pore water conductivity. However, bentonite 1 exhibited excess ions beyond surface adsorption capacity, causing its pore water conductivity to remain higher than bentonite 2 even at increased water content.

\subsection{Experimental setup}

\begin{figure}[t]
	\centering
	\includegraphics[width=0.6\linewidth]{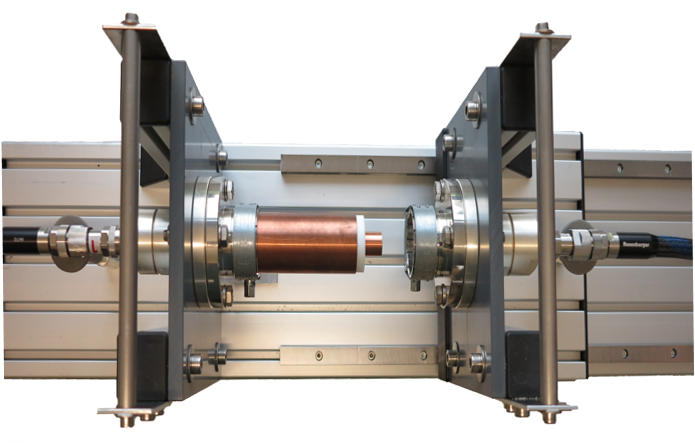}
\caption{Coaxial transmission line cell with adapter units and coaxial cables leading to the Vector Network Analyzer ports. The material under test, here a white PTFE part, is placed as a dielectric between the copper-made inner conductor and the outer conductor of the cell.}
	\label{fig:coax_cell_photo}
\end{figure}

In this investigation, the coaxial transmission line, as shown in Fig. \ref{fig:coax_cell_photo}, served directly as the sample holder \cite{lauer2012new, Wagner2013, bumberger2018, Schmidt2021}. Using adapter units and coaxial cables, the transmission line was connected to the Vector Network Analyzer (VNA) (E5071C, Keysight Technologies Inc., Santa Rosa, CA, US). The VNA measured the scattering parameters ($S_{ij}$) from \SI{300}{MHz} to \SI{3}{GHz} and \SI{5}{GHz}, characterizing the reflection and transmission behavior of the material under test.\\

To ensure accuracy, the influences of the cables and the analyzer were eliminated through a standard calibration with an electronic calibration kit (ECal N4431B, Keysight Technologies Inc., Santa Rosa, CA, US). The N-male to 1 5/8" adapter units (SPINNER GmbH, Munich, Germany) introduced an error in the form of a phase shift. This phase shift was determined using a short-standard in place of the measurement cell and subsequently removed by HF de-embedding \cite{deembed_keysight, bumberger2018}. The dielectric permittivity spectra were calculated from the $S$-parameters using the Backer Jarvis Iterative (BJI) algorithm as described in \cite{lauer2012new, Wagner2013, schwing2016radio}.

\subsection{Inverse parameter estimation \label{sec:Inversion}}

To go beyond simple real and imaginary part models, the proposed models were used to investigate the underlying relaxation mechanisms in the materials. The full set of four scattering parameters was parameterized using a shuffled complex evolution metropolis algorithm (SCEM-UA) \cite{vrugt2003}. This global optimization algorithm combines the strengths of the Metropolis algorithm, controlled random search, competitive evolution, and complex shuffling. It efficiently estimates the optimal parameter set and its underlying posterior distribution within a single optimization run \cite{heimovaara2004}.\\

For the CRIM model, most parameters were determined directly through measurement or assumed based on established properties. However, the pore water conductivity $\sigma_{\mathrm{W}}$ was estimated using a differential evolution algorithm \cite{storn1997}. This approach allowed for a robust and systematic estimation of material parameters, enabling deeper insights into the dielectric properties and relaxation processes of the clay materials.


\section{Results}

\subsection{General Observations of Permittivity Spectra}

\begin{figure}[htb]
	\centering
	\includegraphics[]{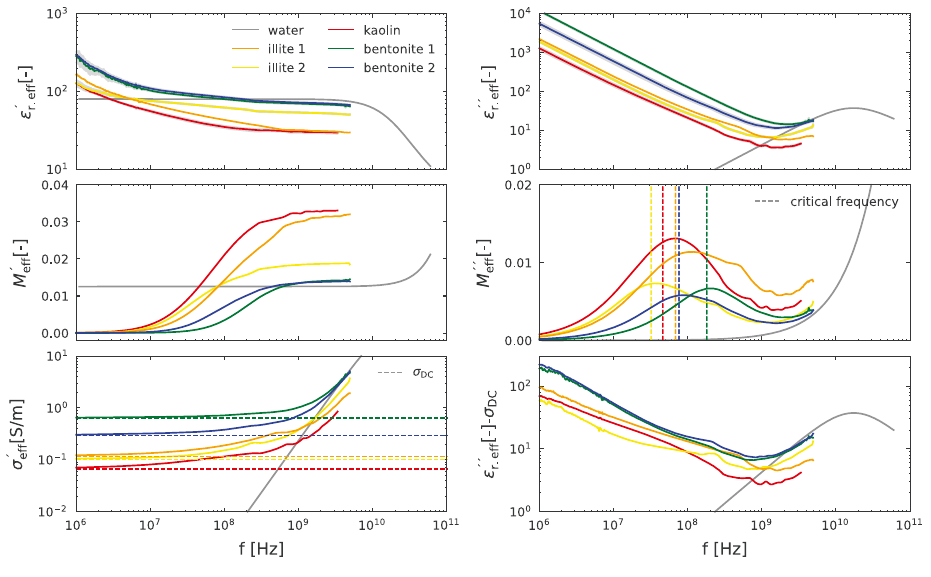}
	\caption{Permittivity spectra of all four clays, together with the spectra of water according to equation \ref{eq:Debye} with values from \cite{kaatze2007} at 21\textdegree C and $\sigma_{\mathrm{w}} = 0$. For this plot, the mean values and standard deviation of the permittivity at each frequency point were determined for each clay type (kaolin, illite, bentonite 1, and bentonite 2). The standard deviation is shown as a grey area. At the critical frequency ($\varepsilon^{''} / \varepsilon^{'} = 1$), the dielectric behavior shifts from conductor to dielectric. The modulus $M^\star$ and effective electrical conductivity $\sigma^\star_{\mathrm{eff}}$ are defined as in Figure \ref{fig:SpectraIntro}. In the lower right plot, the influence of the DC conductivity on the imaginary part has been removed to make other relaxation processes visible.}
	\label{fig:perm_spectra}
\end{figure}

The measured permittivity spectra for all clay samples are presented in Figure \ref{fig:perm_spectra}. The frequency dependence of the real part of the permittivity for the two expansive clays is nearly identical, indicating similar dispersion. However, the imaginary part shows marked differences below 2~GHz, reflecting a stronger DC-contribution for bentonite 1 without affecting relaxation strength.\\

In the lower frequency range below 10~MHz, the real parts exhibit permittivity values that far exceed those of the individual phases (water, clay, air), indicating strong dispersion across all clay samples. This trend flattens at higher frequencies, approaching the GHz-limit of the effective relative permittivity. Meanwhile, in the lower frequency range ($f\leq 100~\mathrm{MHz}$), the typical $1/f$-dependency of the conductivity contribution is evident in the imaginary parts. Around 1~GHz, dielectric losses due to pore water relaxation dominate, causing an increase in the imaginary part (i.e., losses).\\

Modulus representation (Figure \ref{fig:perm_spectra}) reveals significant polarization processes between 10~MHz and 1~GHz, which are obscured in the permittivity plots by low-frequency effects. The peaks of the imaginary part curves align approximately with the critical frequency ($\varepsilon^{''} / \varepsilon^{'} = 1$), which determines the transition between dielectric polarization effects and conductivity effects.\\

\begin{figure}[htb]
	\centering
	\includegraphics[]{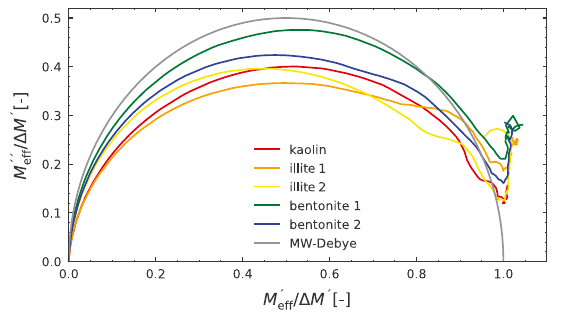}
	\caption{The complex electrical modulus ($M^{\star}_{\mathrm{eff}} = \varepsilon^{\star -1}_{\mathrm{r,eff}}$) calculated from the mean permittivity for each clay material in a Cole-Cole plot. The value of $\Delta M_{\mathrm{MW}} = M_{\mathrm{MW}}^{\prime}(f_{\mathrm{loc.min.}})$ used for normalization is based on the frequency of the local minimum of $M^{\prime\prime}_{\mathrm{eff}}$ around 1~GHz. The MW-Debye curve describes the relaxation behavior of a pure Maxwell-Wagner process, modeled by a Debye function (see equation \ref{eq:HN} and Figure \ref{fig:Empirical}).}
	\label{fig:perm_spectra_modulus_cole}
\end{figure}

The results highlight a strong dependence of dielectric properties on the cation exchange capacity (CEC), water content, and porosity. For example, bentonite 1, with its elevated CEC and porosity due to overactivation, shows unique dielectric dispersion and relaxation strength compared to kaolin and illite. These findings demonstrate the critical role of CEC and porosity in modulating the electromagnetic properties of clay materials (Figure \ref{fig:perm_spectra}).\\

Figure \ref{fig:perm_spectra_modulus_cole} compares the measured data to the pure Maxwell-Wagner process modeled by a Debye-function (see equation \ref{eq:HN} and Figure \ref{fig:Empirical}). The values were normalized to the real part of the modulus at the frequency of the local minimum (around 1~GHz) of the imaginary part; $\Delta M_{\mathrm{MW}} = M_{\mathrm{MW}}^{\prime}(f_{\mathrm{loc.min.}})$. This frequency was determined by smoothing the curve, calculating the first derivative, and identifying its zero crossings. The similarity between the measured curves and the ideal model is evident. However, deviations arise due to superimposed processes, such as physical bound water features and counter-ion effects.\\

\begin{figure}[htb]
	\centering
	\includegraphics[]{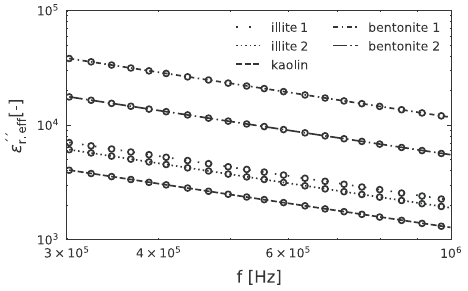}
	\caption{Measurement data (circles) and $\sigma_{\mathrm{DC}}$-fit (lines) of the imaginary part of the effective relative permittivity from 300 kHz to 1~MHz for the four clay materials.}
	\label{fig:sigma_dc_fit}
\end{figure}

The pronounced DC conductivity $\sigma_{\mathrm{DC}}$ of water-saturated clays causes a masking of other relaxation processes in the imaginary part of the effective relative permittivity, extending far into the MHz range. To reveal these masked processes, the influence of DC conductivity must be removed. Using the RF measurement data, the apparent DC conductivity $\sigma_{\mathrm{DC}}^{'}$ can be determined as follows. Assuming that $\sigma^{'}_{\mathrm{eff}} = -\omega \varepsilon_0 \varepsilon^{''}_{\mathrm{r. eff}}$ and that $\sigma^{'}_{\mathrm{eff}}$ is dominated by $\sigma_{\mathrm{DC}}^{'}$ in the frequency range from 300~kHz to 1~MHz, the following function is fitted to the measured data of $\varepsilon^{''}_{\mathrm{r.eff}}$:\\

\begin{equation}
    \varepsilon^{''}_{\mathrm{r.eff}} = \varepsilon_{\infty, \mathrm{app}} -\frac{\sigma_{\mathrm{DC}}^{'}}{\varepsilon_0 \omega}
    \label{eq:sigma_dc_fit}
\end{equation}

The fitting parameters are the apparent RF limit of permittivity $\varepsilon_{\infty ,\mathrm{app}}$ and $\sigma_{\mathrm{DC}}^{'}$. In Figure \ref{fig:sigma_dc_fit}, the measured and fitted data are plotted together to illustrate the suitability of equation \ref{eq:sigma_dc_fit} and its underlying assumptions. The results of this fit are further applied in Figure \ref{fig:perm_spectra} and summarized in Table \ref{tab:sample_data_mean}.\\

The values of porosity and water content vary significantly depending on the dominant clay mineral (see Table \ref{tab:sample_data}). Expansive clays such as bentonite 1 and 2 exhibit higher porosity under saturation conditions compared to non-expansive clays like kaoline and illite. This variation in porosity under water-saturated conditions is reflected in the HF range due to its correlation with the volumetric water content $\theta$. Nonetheless, the analysis enables conclusions regarding the impact of mineralogical and physicochemical properties -- including particle size distribution, specific surface area, clay mineralogy, and cation exchange capacity -- on the dielectric relaxation behavior below 1\,GHz (cf.\ Figure~\ref{fig:perm_spectra}).\\

\begin{figure}[htb]
	\centering
	\includegraphics[width=0.4\linewidth]{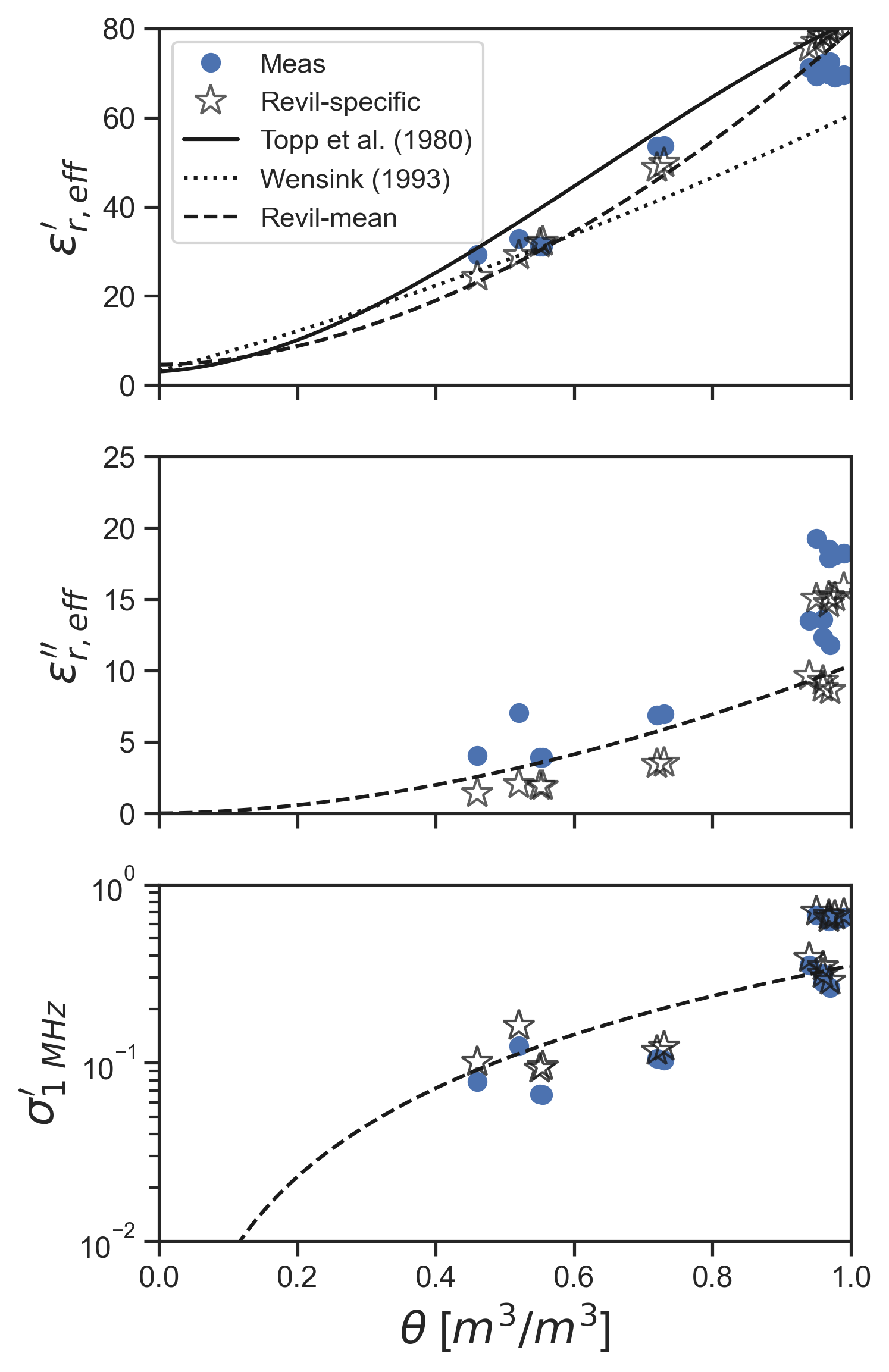}
	\caption{The relative effective permittivity of all samples at $f = 1~\mathrm{GHz}$ as a function of volumetric water content. Additionally, the DC conductivity extrapolated from 1~MHz ($\sigma_{\mathrm{1~MHz}} = \varepsilon^{''}_{\mathrm{r,eff}}(1~\mathrm{MHz}) 2 \pi f \varepsilon_{0}$) is shown in the third plot. The legend applies to all plots. The Topp equation \cite{Topp1980}, the model from Revil et al. \cite{Revil2013} (equations \ref{eq:revil_formula_epsilon} and \ref{eq:revil_formula_sigma}), and the Wensink equation \cite{wensink1993dielectric} (equation \ref{eq:wensink}) are plotted as references. The Revil-specific values correspond to individual samples, while the Revil mean refers to the mean values across all samples for the material parameters in equations \ref{eq:revil_formula_epsilon} and \ref{eq:revil_formula_sigma}.}
	\label{fig1:perm_over_water_cont}
\end{figure}

The dependence of the permittivity at 1~GHz on the volumetric water content is well-established \cite{Topp1980}. Figure \ref{fig1:perm_over_water_cont} illustrates this relationship for the measured samples at 1~GHz. For comparison, the empirical Topp formula (equation \ref{eq:topp}) and the model by Wensink (equation \ref{eq:wensink}) are also plotted. The results show that the real part of the permittivity increases with volumetric water content, eventually approaching the pure water values (e.g., \cite{kaatze2007}) at saturation for high porosity. This behavior aligns with expectations based on the quadratic polynomial nature of equation \ref{eq:topp}. However, for higher water contents or porosities, the Wensink model introduces larger errors in describing the measured results.\\

Since the data presented here aligns with well-known correlations, the validity of the measurements can be confidently assumed. A typical observation is that the real part of the permittivity values remains below those predicted by the Topp formula. This phenomenon is characteristic of clayey soils \cite{wagner2009relationship} and remains a subject of ongoing research \cite{gonzalez2020dielectric}. Investigations by Gonzales et al. \cite{gonzalez2020dielectric} highlight the significant role of porosity and geometrical factors, such as particle shape, orientation, and size distribution, compared to confined or bound water.\\

The imaginary part of the dielectric permittivity characterizes the energy losses associated with electromagnetic wave propagation in the material. These losses comprise both dielectric losses and conduction losses. In the lower frequency range, conduction losses dominate and are primarily determined by the direct current conductivity.\\

Figure \ref{fig1:perm_over_water_cont} also demonstrates that the conductivity of water-saturated samples and bentonite suspensions varies with clay type and water content. Notably, variations in water content within a clay type have a minor influence on conductivity. Furthermore, the conductivities of water-saturated clays correlate positively with the CEC and clay type, , as reported by \cite{bore2022experimental}. Although comparing conductivities across different clay-water ratios introduces potential errors, the trend remains evident. This trend is further supported by examining conductivity values at fixed mixing ratios in Table \ref{tab:sample_data_mean}.\\

A similar trend is observed in the imaginary part $\varepsilon^{''}_{\mathrm{r,eff}}(1~\mathrm{GHz})$, where grouping by clay type and sorting by ascending CEC is evident. While the frequency here is an order of magnitude higher, this behavior suggests that the influence of DC conductivity on the imaginary part extends into the lower GHz range. As polarization losses in water become more prominent, they contribute increasingly to the total losses, reducing the relative influence of clay material-specific losses.\\

\subsection{Direct Dispersion Analysis}

\begin{figure}[htb]
	\centering
	\includegraphics[]{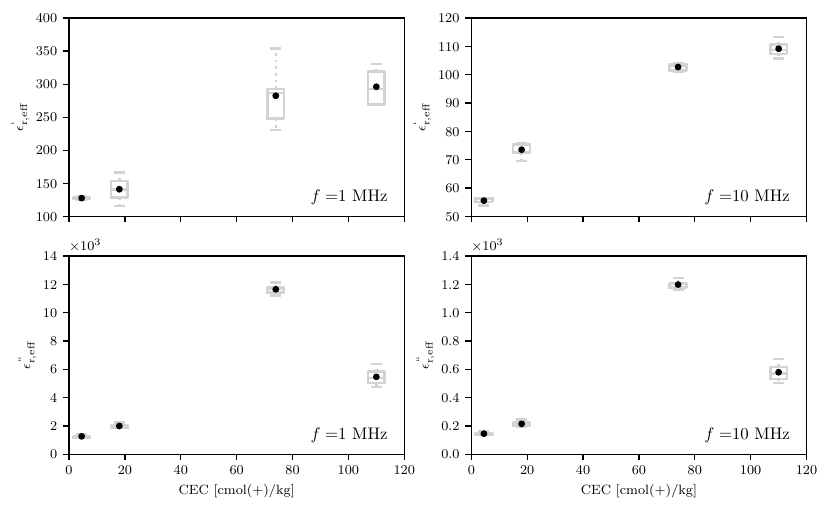}
	\caption{Effective relative permittivity $\varepsilon^\star_{\mathrm{r,eff}}(f)$ of the four clay materials at $f = 1~\mathrm{MHz}$ (left) and $f = 10~\mathrm{MHz}$ (right) as a function of the CEC. The respective boxplots, with additional mean values ($\bullet$), show the statistics of all measurements for each clay type.}
	\label{fig1:josh_plots}
\end{figure}

An approach to analyse the dielectric spectra of clay-containing soils and rocks is to evaluate the spectra at specific frequencies, as described in \cite{Josh2015}. Figure~\ref{fig1:josh_plots} presents the values for $\varepsilon^{*}_{\mathrm{r,eff}}$ plotted against the CEC at $f = 1~\mathrm{MHz}$ and $f = 10~\mathrm{MHz}$ for both the real and imaginary components.\\

At 1 MHz and 10 MHz, the effective relative permittivity demonstrates clear correlations with CEC and water content. Notably, the imaginary part shows a strong positive correlation with CEC, distinguishing clay types effectively (Figure \ref{fig1:josh_plots}). The results also emphasize that higher water content and porosity, as seen in bentonite samples, amplify relaxation processes, particularly at lower frequencies, where conductivity dominates.\\

The real parts of the effective dielectric permittivity are sorted in ascending order according to their CEC. The imaginary part is statistically more robust, as evidenced by the height of the boxplots, and provides valuable insights due to its correlation with CEC \cite{Revil2017complex}. In Figure~\ref{fig1:josh_plots}, the different clay materials are distinctly identifiable by the imaginary part of the permittivity. Furthermore, materials with low CEC exhibit relatively small values of the imaginary part of the effective relative permittivity, consistent with expectations. Notably, the imaginary parts at CEC=74~cmol(+)/kg are higher compared to those at CEC=110~cmol(+)/kg. Both clays are bentonites activated with sodium carbonate, but bentonite 1 received a significantly larger amount of sodium carbonate than bentonite 2. This leads to an oversupply of ions in the suspension that are not adsorbed onto the clay surface, substantially increasing the conductivity of the suspension and disrupting the natural correlation between CEC and conductivity. Consequently, the high values of the imaginary part for bentonite 1 dominate the dielectric spectrum, largely influenced by conductivity.\\

Josh et al. \cite{Josh2015} discussed that electrode polarization is relevant even in permittivity spectra below 10~MHz obtained by capacitance methods. However, this frequency range also includes effects such as Maxwell-Wagner polarization, ionic conduction in the porous network, and restricted ion mobility within the electrical double layer \cite{Revil2013}. The Maxwell-Wagner effect results in permittivities that far exceed those of individual components and depends strongly on material contrast, surface area, and the volume fractions of minerals and fluids \cite{Josh2015}.\\

In the frequency range above 10~MHz, pore geometry effects on dielectric losses become less pronounced. Consequently, spectral analyses at $f >$ 10~MHz are less sensitive to interference. In this range, charge transport paths are shortened, and as a result, the Maxwell-Wagner effect dominates, while double-layer effects have a minor impact on the dielectric relaxation behavior. This shift enhances the influence of clay surface charge effects on the dielectric spectrum. Materials with a larger surface area or higher CEC exhibit greater polarizability, leading to higher real parts of the effective permittivity.\\

This trend is evident when comparing the two frequency points in Figure~\ref{fig1:josh_plots}. The real parts of the permittivities at $f=$10~MHz show a clear positive correlation with CEC and exhibit reduced data spread. In contrast, at $f=$1~MHz, the data spread within each clay type is more substantial, likely due to additional effects influencing polarizability at lower frequencies.\\ 

An additional observation emerges from comparing the imaginary parts at the two frequency points. The imaginary part values at $f=$1~MHz and $f=$10~MHz differ by approximately a factor of 0.1, supporting the assumption of $1/f$-dependence for the imaginary component and the dominance of apparent DC conductivity.\\

\subsection{Decomposition of the Dielectric Spectra}

\begin{figure}[htb]
	\centering
	\includegraphics[width=\textwidth]{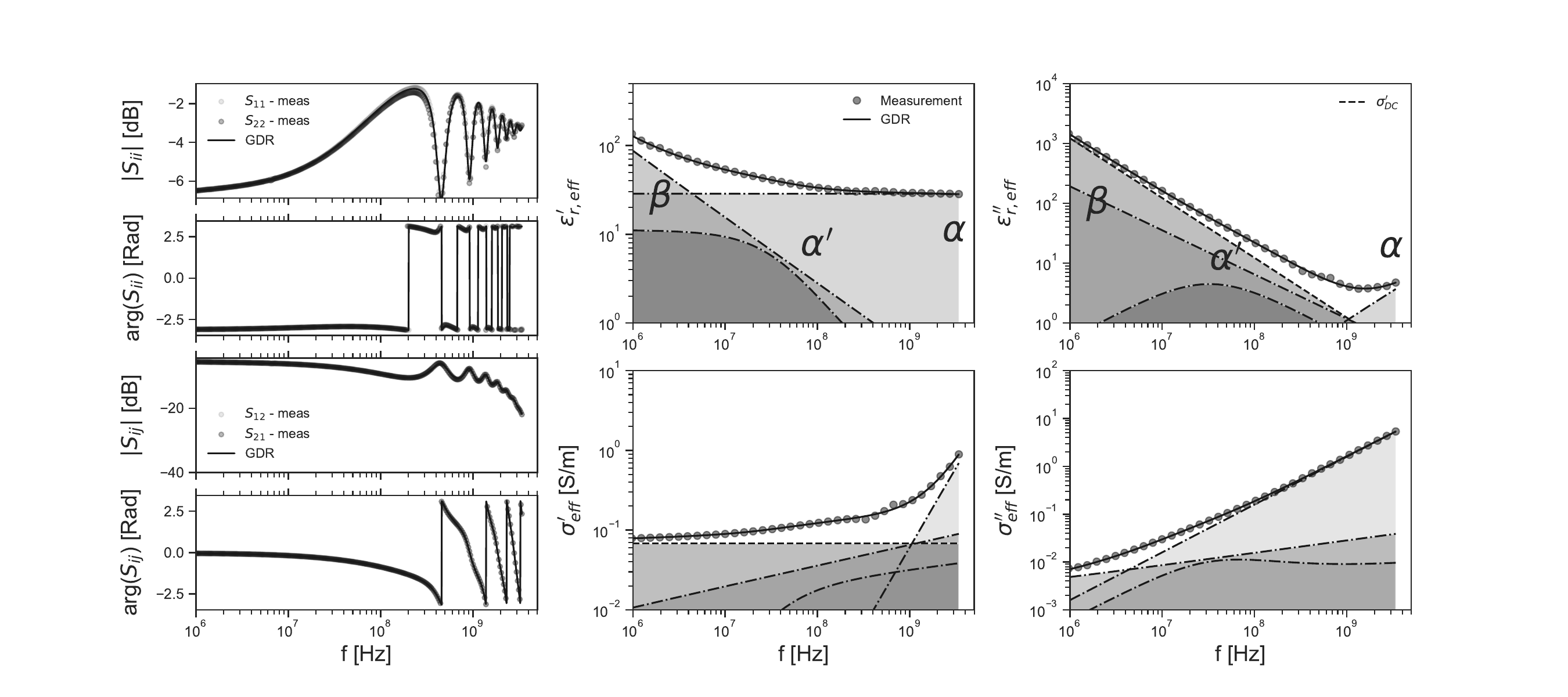}
\caption{Measured and modeled (GDR - generalized dielectric response) HF-EM material properties of kaolin sample ID602 over frequency. Left: S-parameter. Center: complex effective relative permittivity. The HF-water relaxation contribution is depicted by $\alpha$. The other terms are $\alpha'$ for the intermediate and $\beta$ for the low-frequency relaxation processes. Right: complex effective electrical conductivity.}
	\label{fig:gdr_plots}
\end{figure}

The results of the inverse parametrization of the measured S-parameters using the described models are shown in Figures~\ref{fig:gdr_plots}, \ref{fig:cpcm_plots}, \ref{fig:crim_plots}, and \ref{fig:abc_plots}. Except for the CRIM, the models accurately describe the dielectric responses of the clays. The parametrizations effectively reproduce the spectral curves, as depicted in the figures. To highlight the differences in the modeling approaches, the individual relaxation contributions are also included in the diagrams.\\

\begin{figure}[htb]
	\centering
	\includegraphics[]{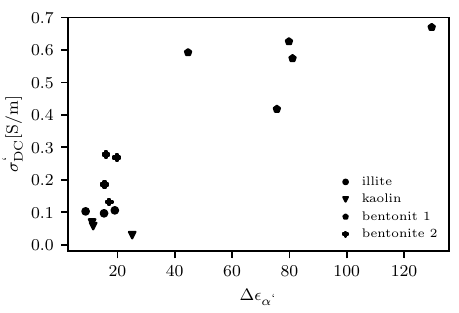}
\caption{Relaxation strength of the second relaxation process versus direct current conductivity $\sigma'_{\mathrm{DC}}$ in the GDR model.}
	\label{fig1:dispersion}
\end{figure}

For the GDR model, the relaxation contributions include RF-water relaxation depicted as $\alpha$, intermediate relaxation as $\alpha'$, and low-frequency relaxation as $\beta$. In the kaolin example (Figure~\ref{fig:gdr_plots}), these components are distinctly identifiable. Comparing results for the different clays (Figure~\ref{fig1:dispersion}) reveals that for illite, kaolin, and bentonite 2, the relaxation strength of the $\alpha'$ process remains relatively consistent, with values ranging from 8.9 to 25.2. In contrast, bentonite 1 exhibits significantly higher relaxation strength, varying from 44.7 to 129.7, reflecting the overactivation with sodium carbonate in this sample.\\

The CPCM is particularly noteworthy because it incorporates low-frequency conductivity parameters, such as the apparent chargeability $M'$, to model the lower frequency range (see Equation~\ref{eq:cpcm_summe}). As described by Revil et al. \cite{Revil2017complex}, the normalized chargeability $M^{`}_{\mathrm{n}} = M'\sigma_{\infty} = \sigma_{\infty} - \sigma_0$ provides insight into the correlation between relaxation parameters and soil properties. However, since $\sigma_{\infty}$ is not directly available as a parameter in the model, $M^{`}_{\mathrm{n}}$ is calculated as $\sigma_0 M' / (1-M)$.\\

The GDR model confirms the dependence of relaxation strength on CEC, water content, and porosity. For instance, the $\alpha'$ relaxation strength of bentonite 1 is significantly higher (44.7–129.7) than that of kaolin and illite (8.9–25.2), correlating with its high CEC and porosity. This demonstrates the influence of CEC on low-frequency dispersion processes (Figure \ref{fig1:dispersion}).\\

\begin{figure}[htb]
	\centering
	\includegraphics[width=\textwidth]{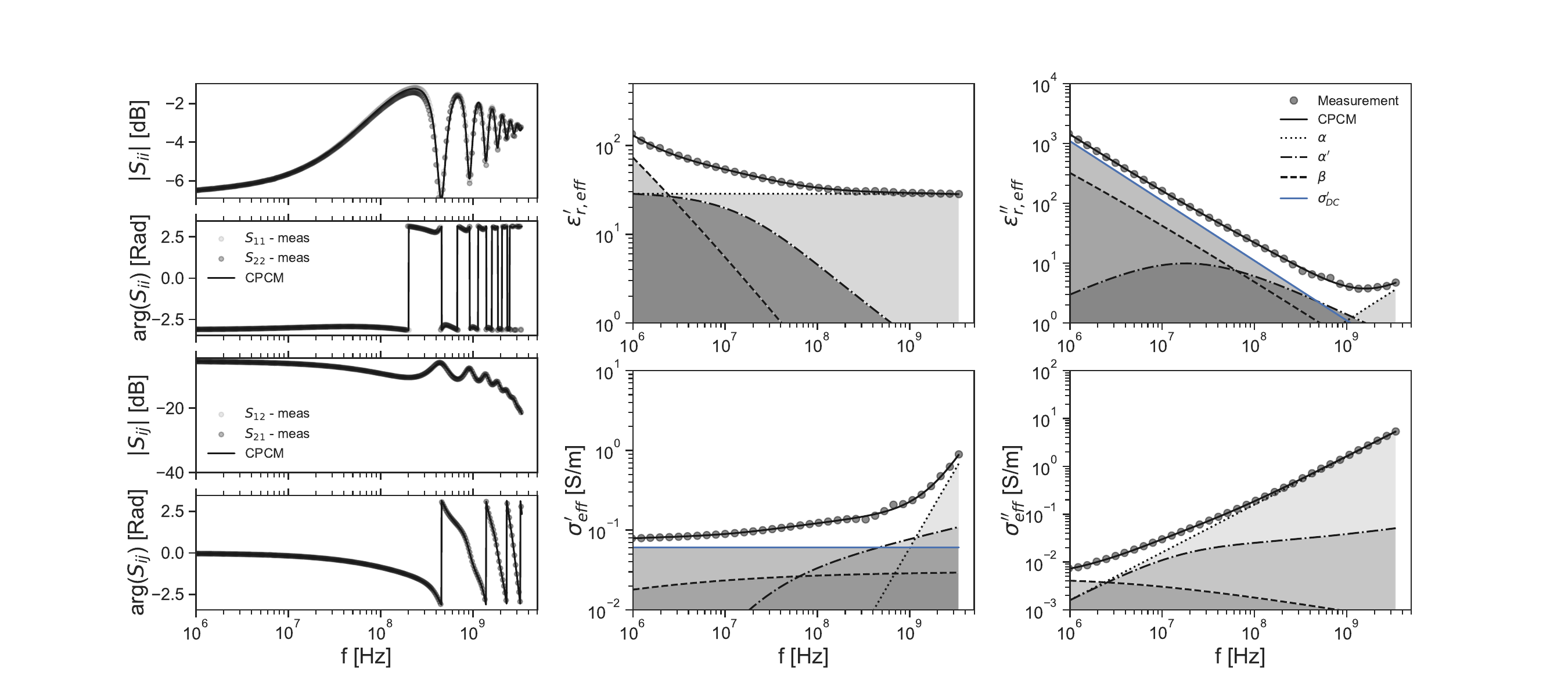}
	\caption{Measured and modeled CPCM (Combined Permittivity and Conductivity Model) HF-EM material properties of kaolin sample ID602 over frequency. Left: S-parameter. Center: complex effective relative permittivity.}
	\label{fig:cpcm_plots}
\end{figure}

\begin{figure}[htb]
	\centering
	\includegraphics[]{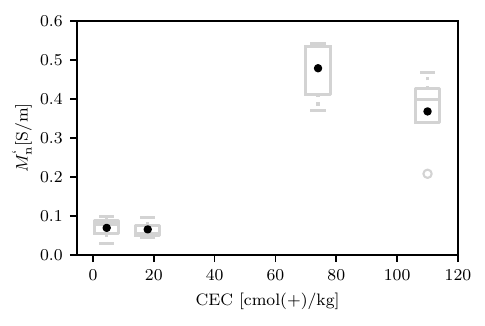}
	\caption{Normalized apparent chargeability $M^{\prime}_{\mathrm{n}}$ of the third relaxation process in the Combined Permittivity and Conductivity Model (CPCM) as a function of the CEC.}
	\label{fig1:chargeability_over_cec}
\end{figure}

Figure~\ref{fig1:chargeability_over_cec} illustrates $M^{`}_{\mathrm{n}}$ plotted against the CEC to explore the relationship between chargeability and CEC. According to the model proposed by Revil et al. \cite{Revil2017complex}, chargeability is expected to positively correlate with CEC, assuming that relaxation processes in this frequency range predominantly originate from double-layer interactions.\\

However, since different relaxation effects overlap in the middle and lower frequency ranges, and the apparent chargeability is an extrapolated parameter, a clear dependence of chargeability on the CEC cannot be established unequivocally. Consequently, no definitive correlation is observed in Figure~\ref{fig1:chargeability_over_cec}.\\

\subsection{Mixture Models}

\begin{figure}[htb]
	\centering
	\includegraphics[width=0.85\textwidth]{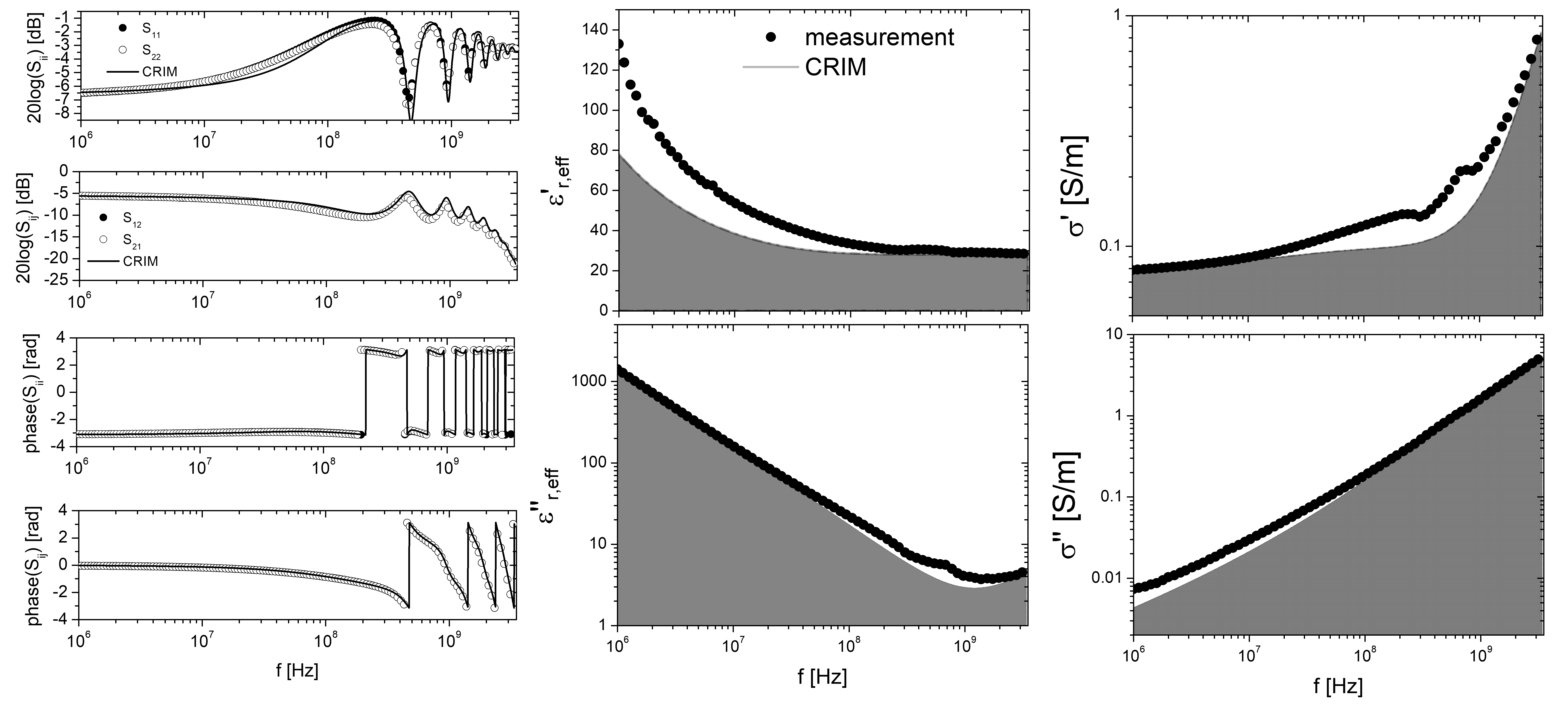}
	\caption{Measured and modeled (CRIM - Complex Refractive Index Model) HF-EM material properties of kaolin sample ID602 over frequency. Left: S-parameter. Center: complex effective relative permittivity. Right: complex effective electrical conductivity.}
	\label{fig:crim_plots}
\end{figure}

The spectra of soils with high clay content can only partially be described by the CRIM mixture approach, as shown in Figure~\ref{fig:crim_plots}. Above \SI{400}{MHz}, the model provides a good estimation of the amount of free pore water by accurately modeling $\varepsilon'_{\mathrm{r,eff}}$. Although the direct current conductivity aligns reasonably well with the measurements, the CRIM approach fails to adequately capture low-frequency dispersion, limiting its applicability for characterizing dielectric behavior in this range.\\

\begin{figure}[htb]
	\centering
	\includegraphics[width=\textwidth]{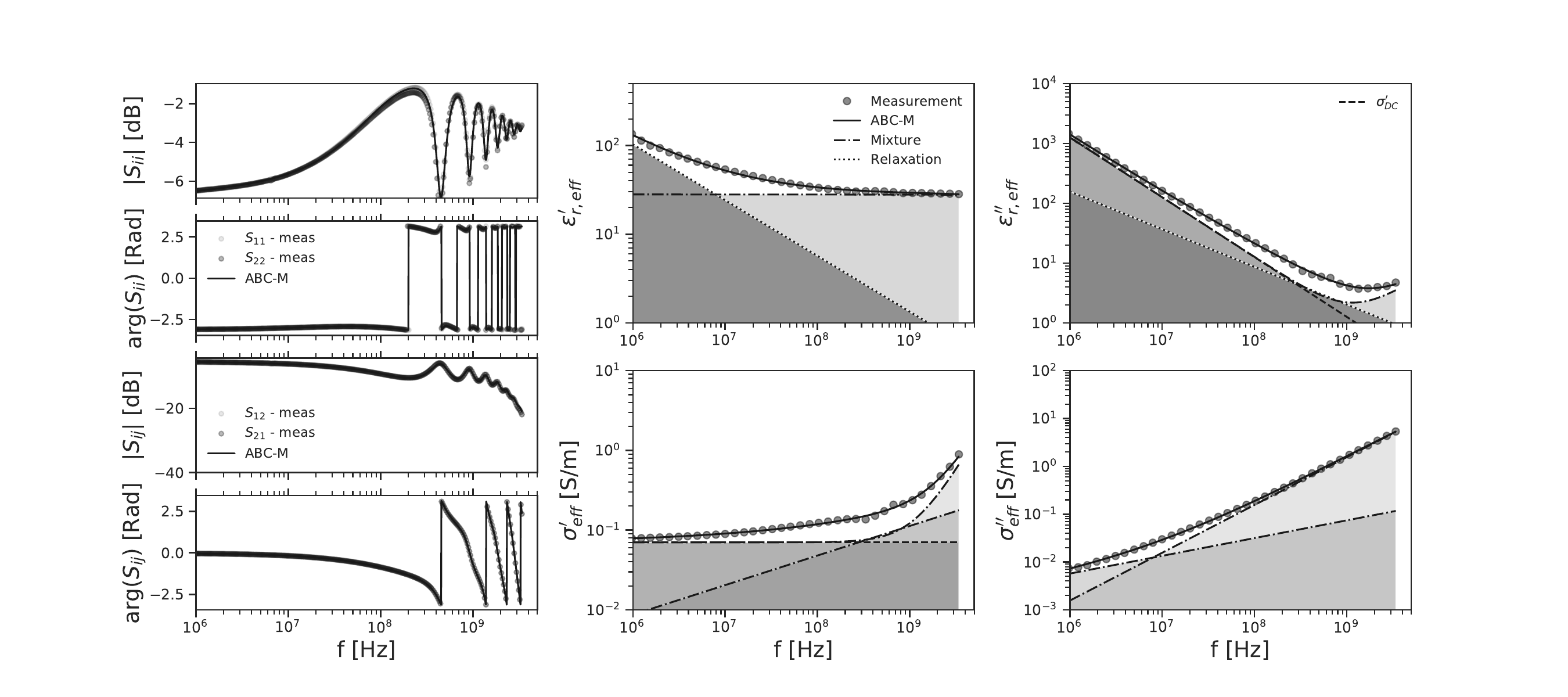}
	\caption{Measured and modeled (ABC-M - Augmented Broadband Complex Dielectric Mixture Model) HF-EM material properties of kaolin sample ID602 over frequency. Left: S-parameter. Center: complex effective relative permittivity. Right: complex effective electrical conductivity.}
	\label{fig:abc_plots}
\end{figure}

The ABC-M approach (Figure~\ref{fig:abc_plots}) demonstrates greater flexibility for characterizing the full spectrum due to the inclusion of an additional relaxation term. This flexibility allows the ABC-M to better represent both high- and low-frequency behavior compared to the CRIM approach. However, the physical interpretation of the exponents $m$ and $n$ in the mixture term (see Figure~\ref{fig:parameter_all_models}) requires further investigation. These parameters are highly sensitive to variations in water saturation and porosity, which strongly influence the relationship between volumetric water content $\theta$ and the relative permittivity at high frequencies, particularly $\varepsilon_{\mathrm{1~GHz}}$.\\

The CRIM model effectively captures high-frequency behavior above 400 MHz, where water content and porosity dominate the dielectric response. However, it underrepresents the influence of CEC and associated low-frequency dispersion effects, highlighting the importance of more advanced models for clays with high CEC and porosity (Figure \ref{fig:crim_plots}).\\


\section{Discussion}

The GDR, CPCM, and ABC-M models successfully replicate the broadband relaxation behavior of water-saturated clays over the investigated frequency range (Figures \ref{fig:gdr_plots}, \ref{fig:cpcm_plots}, \ref{fig:crim_plots}, \ref{fig:abc_plots}). In contrast, the CRIM model accurately fits the high-frequency (HF, >100 MHz) real part of the complex effective permittivity and the low-frequency (LF, $f \leq$ 100 MHz) imaginary part, but it does not adequately model dispersion phenomena. CPCM and GDR are purely phenomenological relaxation models, allowing for the comparison of individual relaxation processes. Meanwhile, ABC-M combines the strengths of a mixture equation with the adaptability of a relaxation model, enabling the decomposition of dispersion into distinct superimposed relaxation phenomena, which can be attributed to different physical causes.\\

These results align with the central hypothesis of this study, demonstrating that cation exchange capacity (CEC), water content, and porosity are significant factors influencing the dielectric properties of water-saturated clays. The modeling approaches and experimental analyses confirm the relevance of these physicochemical parameters, emphasizing the validity of the chosen methodology.\\

\begin{figure}[htb!]
	\centering
		\includegraphics[]{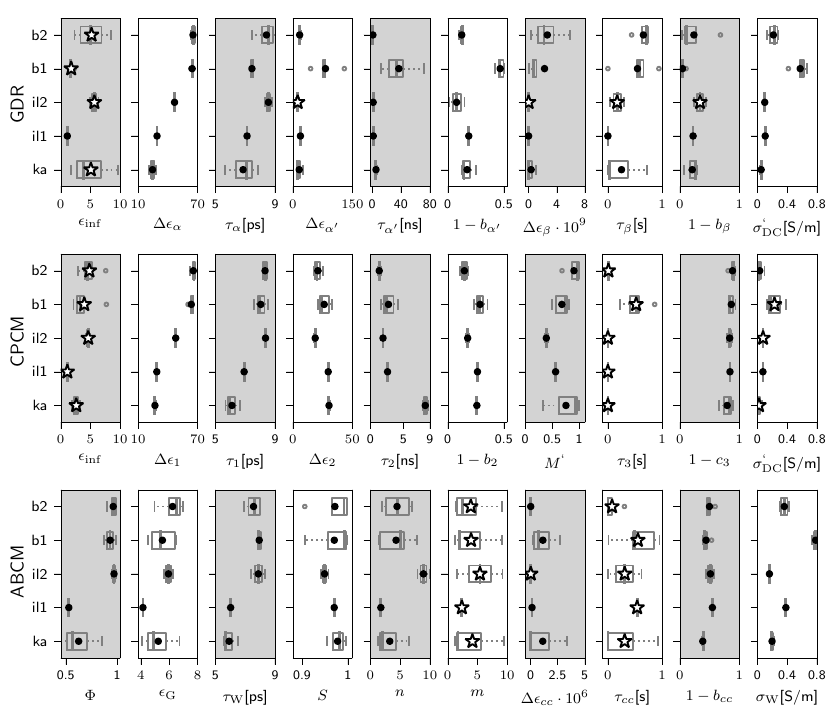}
	\caption{Results of the inverse parameter estimation for three models: Fractional Dielectric Relaxation Model (GDR), Combined Permittivity and Conductivity Model (CPCM), and Augmented Broadband Complex Dielectric Mixture Model (ABC-M). For each clay type (ka - kaolin, il1 - illite 1, il2 - illite 2, b1 - bentonite 1, b2 - bentonite 2), the set statistics are represented by a boxplot for each parameter individually. Additionally, the mean is depicted by a $\bullet$; or by a $\openbigstar[.6]$ if the inversion parameters have an uncertainty greater than 10\% of the parameter value.}
	\label{fig:parameter_all_models}
\end{figure}

A detailed comparison of the inversion results (Fig. \ref{fig:parameter_all_models}) reveals that the models produce partially consistent results in frequency ranges dominated by specific relaxation phenomena, such as water relaxation.\\

\subsection{Water Relaxation}
One parameter consistent across all three models is the position of the free water relaxation peak, which is precisely determined despite the restricted frequency window with a maximum of 5 GHz. For all models and clay types, the relaxation frequency is observed between 5 and \SI{9e-12}{\second} (200 -- 111~GHz), aligning well with values reported in \cite{kaatze2007}. The mean value for kaolin, 6.3~ps (159~GHz), is slightly lower than the other clays, which average 7.9~ps (127~GHz).\\

In the CPCM and GDR models, additional parameters of the water Debye term, $\varepsilon_{\mathrm{inf}}$ and $\Delta \varepsilon_{1,\alpha}$, are also inverted. Both models yield nearly identical values for the high-frequency limit ($\approx 5$), consistent with literature values for pure water \cite{kaatze2007}. Furthermore, the relaxation strength demonstrates a positive correlation with the volumetric water content, as expected. A higher volumetric water content contributes more significantly to the overall polarizability.\\

\subsection{Intermediate Relaxation}
The intermediate relaxation process, situated between the RF water relaxation and the LF process, is modeled in both the GDR and CPCM approaches using a Cole-Cole term. Therefore, the results are expected to be at least partially comparable. Regarding the relaxation peaks $\tau_{2,\alpha'}$, the CPCM estimates range from \SI{1.1}{\nano\second} to \SI{8.8}{\nano\second} (\SI{114}{\MHz} - \SI{909}{\MHz}), while the GDR values span from \SI{1.1}{\nano\second} to \SI{71}{\nano\second} (\SI{14}{\MHz} - \SI{909}{\MHz}). All observed values fall within the MHz range. The CPCM inversion values for kaolin stand out, with an average relaxation time of \SI{8.3}{\nano\second} (\SI{19}{\MHz}), significantly higher than the average of other clays at \SI{2.1}{\nano\second} (\SI{76}{\MHz}). This deviation warrants further investigation, particularly concerning the microscopic clay structure.\\

The relaxation strengths, $\Delta \varepsilon_{2}$, estimated by the CPCM range from $18 \leq \Delta \varepsilon_{2} \leq 33$, which aligns closely with the GDR estimates of $9 \leq \Delta \varepsilon_{\alpha'} \leq 25$. A slight positive correlation between relaxation strength and relaxation time is observed, largely attributed to the pronounced $1/\log (f)$ dependence of the imaginary part $\varepsilon^{''}_{\mathrm{r,eff}}$ in this frequency range. Interestingly, the GDR results for bentonite 1 show substantially higher relaxation strengths of $45 \leq \Delta \varepsilon_{\alpha'} \leq 130$, likely due to the stronger soda activation.\\

The observed relationship between CEC and relaxation strength, particularly for bentonite 1, underscores the critical role of CEC in determining low-frequency dispersion processes. This finding highlights the potential of these models to serve as diagnostic tools for quantifying CEC indirectly through dielectric measurements.\\

\subsection{Low-Frequency Relaxation}
In the lower frequency range, the relaxation process dominates the overall frequency dependence of $\varepsilon'_{\mathrm{r,eff}}$. Therefore, any broadband relaxation model or mixture equation must incorporate this process to accurately reflect its significant contribution. The approaches used here-GDR, CPCM, and ABC-M-each include a relaxation term (referred to as the third process) to model this behavior. In contrast, CRIM accounts for this dispersion implicitly through the structural exponent 0.5 \cite{wagner2011}. While GDR and ABC-M use formulations within the permittivity domain, CPCM employs a conductivity representation. All three models (GDR, CPCM, and ABC-M) accurately predict the low-frequency dispersion, whereas in CRIM, the structural exponent must be adjusted to reflect particle size distribution and particle orientation in the applied electrical field \cite{wagner2011, Sihvola2000, Jones2000a}. These differing modeling domains lead to variations in parameter estimates for describing the dispersion.\\

While the models effectively capture broadband dielectric behavior, certain limitations remain. For instance, the CRIM model struggles with low-frequency dispersion, and the ABC-M model requires further refinement to decouple correlated parameters like porosity and solid particle permittivity. Future work could focus on extending the frequency range of measurements or incorporating additional physical constraints to enhance model accuracy.\\

For parameters such as $1-b_{\beta}$, $1-c_3$, and $1-b_{\mathrm{cc}}$, which define the width and shape of the relaxation peak, the results are consistent within each model but exhibit significant differences across models. Similarly, the position of the relaxation peak on the frequency axis, characterized by the relaxation time $\tau$, does not yield consistently similar values across models. Notably, a strong parameter uncertainty, highlighted by $\openbigstar[.6]$, is observed. This is expected, as the relaxation frequencies and magnitudes of the assumed processes lie outside the measured frequency window.\\

This study contributes to the growing body of knowledge on dielectric spectroscopy by providing a comprehensive analysis of multiple modeling approaches. By correlating dielectric parameters with CEC, water content, and porosity, the findings bridge experimental observations with theoretical predictions, offering a robust framework for future research in soil and material sciences.\\

Furthermore, the relaxation phenomena in the lower frequency range---such as double-layer relaxation, counter-ion relaxation, and the Maxwell-Wagner effect---exhibit a strong dependence on geometry and are often superimposed. The complex pore and particle shapes result in numerous possible frequency dependencies, leading to a broad and smooth relaxation signature \cite{wagner2011}. Relaxation strengths, $\Delta \varepsilon_{\beta}$, $M'$, and $\Delta \varepsilon_{\mathrm{cc}}$, vary significantly across models, underscoring the influence of model formulations. This suggests that combining theoretical mixture equations with flexible relaxation terms, such as Cole-Cole-type functions, could more effectively connect the HF and LF ranges.\\

The ability of the models to accurately represent relaxation processes has both scientific and practical implications. For example, these findings can be applied in geotechnical engineering and soil science to better evaluate the mechanical and hydraulic properties of soils. Moreover, these models could assist in optimizing industrial processes that rely on the understanding of clay-water interactions, such as drilling fluid design or soil stabilization techniques.\\

\subsection{DC Conductivities}
The conductivity values, shown on the far-right side of Figure \ref{fig:parameter_all_models}, are in a similar range across all three models. However, the pore water conductivity is somewhat higher for ABC-M compared to CRIM. This difference is plausible, as the pore water conductivity in ABC-M is treated separately, while the apparent DC conductivities also account for surface conductivity contributions, as described by the modified Archie model \cite{Revil2007}:

\begin{equation}\label{eq:Archie}
\sigma_{\mathrm{eff}}^\star = \frac{\sigma_{w}^\star}{F} + \left(1 - \frac{1}{F}\right)\sigma_{S}^\star,
\end{equation}

where $\sigma_{w}^\star$ is the complex conductivity of the aqueous pore solution, $\sigma_{S}^\star$ is the complex surface conductivity, and $F=\phi^{-m}$ is the formation factor, which relates to porosity $\phi$ using the cementation exponent $m$. The complex pore water conductivity, $\sigma_{w}^\star$, is modeled using the approach by \cite{Jougnot2010}:

\begin{equation}\label{eq:Jougnot}
\sigma_{w}^\star = j\omega\varepsilon_0\varepsilon_{r,W}^\star,
\end{equation}

where $\varepsilon_{r,W}^\star$ is the relative complex permittivity of the aqueous pore solution, derived from equation (\ref{eq:Debye}), including $\sigma_w$. As discussed earlier, bentonite 1 deviates from these conductivity values due to its strong sodium carbonate activation.\\

A noticeable difference is observed in the $\sigma'_{\mathrm{DC}}$ values between the GDR and CPCM models. The CPCM-derived values are systematically lower than those of GDR. This discrepancy arises because these conductivities are apparent DC conductivities extracted from RF measurements rather than measured directly. Additionally, in CPCM, $\sigma'_{\mathrm{DC}}$ becomes frequency-dependent through the third relaxation term (see equation \ref{eq:cpcm_summe}), resulting in slightly lower values. The relaxation frequency of the conductivity process in CPCM is constrained to frequencies above 10 MHz, making the model more suitable for measurements that include spectroscopic data from the kHz range. However, the apparent chargeability $M'=\sum_{k}^N m_k$ obtained in CPCM, serves as a measure of the total polarization contribution, encompassing all superimposed processes.\\

The apparent conductivity of the aqueous pore solution, $\sigma_{\mathrm{W}}$, derived from CRIM (Table \ref{tab:sample_data}) and ABC-M, as well as the apparent DC conductivities, $\sigma'_{\mathrm{DC}}$, estimated by GDR and CPCM, exhibit a strong correlation with CEC. The correlation between DC conductivity and CEC observed across the models further emphasizes the relevance of CEC as a diagnostic parameter in soil science. By linking DC conductivity with clay-water interactions, this study highlights its potential as a proxy for assessing soil properties such as permeability and ionic mobility. This connection strengthens the practical applicability of the presented approaches for real-world geotechnical and environmental challenges.\\

Once again, bentonite 1 deviates from the expected relationship due to its high sodium carbonate activation. As CEC increases, more cations are present on the inner surfaces of clay mineral particles. These samples were saturated with deionized water, leading to higher concentrations of free charges in the aqueous pore solution and systematic variations in conductivity across all models. However, in bentonite 1, the excessive free ions in the suspension, which are not solely dissociated from the clay surface, disrupt this correlation. The appropriate volume fraction of water, as indicated in Table \ref{tab:sample_data}, mitigates this increase in ion concentration, reducing its impact on relaxation parameters, including direct current conductivity.\\

\subsection{ABC-M Specific Parameters}
\begin{figure}[htb!]
    \begin{center}
				\includegraphics[]{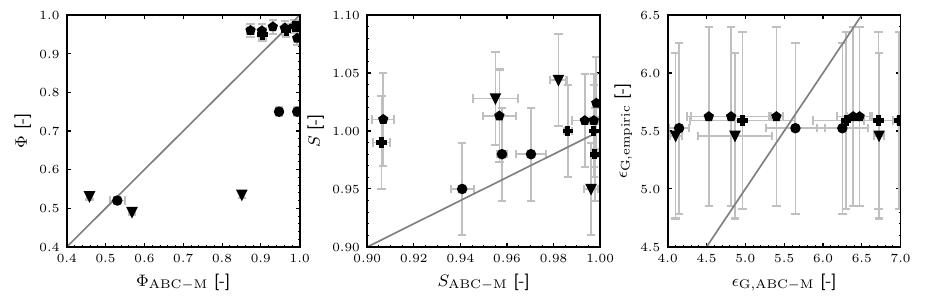}
    \end{center}
    \caption{Comparison of ABC-M model-inverted parameters: porosity $\phi$, water saturation $S$, and permittivity of the solid particles $\varepsilon_{\mathrm{G}}$ with measurements by other methods. Grey uncertainty bars are included, and the diagonal line represents a 1:1 correlation.}
    \label{fig:abcm_compare}
\end{figure}

Due to the formulation of the ABC-M mixture model, parameters such as porosity $\phi$, permittivity of the solid particles $\varepsilon_{\mathrm{G}}$, and saturation $S$ can be estimated from the measured spectra by comparing them to the determined parameters (see Section 3). Unlike the CRIM inversion, where certain parameters are pre-determined, the ABC-M model leaves all parameters open, even when some are known. This flexibility allows for an assessment of the model's performance when analyzing completely unknown materials. The comparison of these measurements is shown in Figure \ref{fig:abcm_compare}.\\

As observed in Figure \ref{fig:abcm_compare}, the majority of the porosity values are well estimated by the ABC-M model. However, the presence of three outliers suggests a potential correlation with other parameters. Saturation values, which are close to 1, are also well predicted by the model. In the case of the permittivity of the solid particles $\varepsilon_{\mathrm{G}}$, the values determined by the helium pycnometer and the formulas (\ref{eq:olhoeft}) and (\ref{eq:Dobson}) range between 4.5 and 5.0. However, the values estimated by ABC-M vary between 4 and 7, indicating that the model struggles to determine this parameter accurately. This discrepancy likely arises because $\varepsilon_{\mathrm{G}}$ is correlated with porosity and the cementation exponent $m$ in the model. Consequently, incorporating a priori information to fix this parameter would improve the model's reliability.\\

The saturation exponent $n$ and the cementation exponent $m$ exhibit significant variation within their inversion bounds. For a single clay type, individual sample values often vary widely, with illite spanning nearly the entire boundary interval. This behavior is even more pronounced for $m$, which shows similar distributions across all four clay types. Such variability is expected since $S_W \approx 1$, rendering the saturation exponent $n$ almost irrelevant to the modeled effective permittivity.\\

The ABC-M model's ability to estimate porosity and water saturation aligns well with the study's goal of linking dielectric behavior to fundamental soil parameters. However, the challenges in accurately determining the permittivity of solid particles highlight the need for incorporating external constraints or more precise parameterizations in future work. These refinements could enhance the robustness of ABC-M for broader applications in soil characterization and material design.\\

In cases of near saturation, the ABC-M mixture term must be reduced to the following form, as described in \cite{bore2018}:

\begin{equation}
    \varepsilon_{\mathrm{r, mix}}^\star = \phi^m[\varepsilon_{\mathrm{r,W}}^\star +
    (\phi^{-m}-1)\varepsilon_{\mathrm{G}}],
\end{equation}

where $\varepsilon_{\mathrm{r,W}}^\star$ represents the complex relative permittivity of water and $\varepsilon_{\mathrm{G}}$ corresponds to the permittivity of the solid particles. This reduced form highlights the diminishing influence of $n$ under saturated conditions and emphasizes the need for accurate parameterization of $\phi$, $\varepsilon_{\mathrm{G}}$, and $m$.


\section{Conclusion}
In this study, the dielectric spectra of four different water-saturated clays were analyzed in the frequency range from 1~MHz to 5~GHz. The investigation employed direct spectrum analysis, semi-empirical and theoretical mixture models, as well as phenomenological relaxation functions to explore the relationships between dielectric properties and clay-specific parameters.\\

One central hypothesis of this study was that the mineralogical composition of water-saturated clays significantly influences their high-frequency electromagnetic response. This, in turn, would allow material-specific parameters, such as the cation exchange capacity (CEC), to be extracted from broadband dielectric spectra. The results indicate that this hypothesis holds qualitatively true for most cases, with some notable exceptions.\\

By directly analyzing permittivity and conductivity spectra, qualitative correlations with clay-specific parameters were observed. For instance, the real parts of the spectra at 10~MHz demonstrated a positive correlation with CEC. The influence of the specific surface area, particularly the electrostatic double layer, was found to dominate over geometric effects, such as the Maxwell-Wagner polarization. This supports the observation that permittivity correlates more strongly with CEC and specific surface area. On the other hand, the imaginary part at 10~MHz was shown to depend heavily on the conductivity of the material.\\

These observations underscore the importance of clay-specific surface properties, such as the electrostatic double layer, in influencing dielectric parameters. The differentiation of relaxation phenomena also highlights the challenges of disentangling overlapping processes in complex geomaterials, necessitating further refinement of analytical and modeling techniques.\\

In addition, the findings emphasize the relevance of dielectric spectroscopy as a non-invasive tool to link electromagnetic responses to specific properties such as CEC, porosity, and water content. This study advances the understanding of how mineralogical composition and ion exchange processes in clays govern their broadband dielectric behavior, providing a foundation for practical subsurface characterization methods in geophysical and geotechnical applications.\\

This behavior was consistent across naturally occurring clays like kaolin, illite, and bentonite 2. However, the strongly soda-activated bentonite 1 deviated from expectations, demonstrating that chemical treatments significantly impact the dielectric spectra.\\

Quantitative estimations using adjusted models such as ABC-M and CPCM proved challenging due to overlapping relaxation phenomena within this frequency range, which could not be fully disentangled. Additionally, models that integrate dielectric properties with chemical parameters, such as CEC, rely on low-frequency measurements, resulting in inaccuracies when extrapolated to the high-frequency range.\\

By integrating experimental data with advanced modeling approaches such as GDR, CPCM, and ABC-M, this research contributes to bridging the gap between fundamental mineral physics and applied electromagnetic methods. The ability of these models to predict material properties with high accuracy highlights their potential for broader applications in field-scale geophysical monitoring.\\

The three models---Generalized Dielectric Relaxation Model (GDR), Combined Permittivity and Conductivity Model (CPCM), and Augmented Broadband Complex Dielectric Mixture Model (ABC-M)---successfully predicted the broadband dielectric spectra of the materials with high accuracy. However, the Complex Refractive Index Model (CRIM) was less effective at accounting for strong relaxation processes below 400~MHz. Furthermore, this study confirmed that water content determination of clayey geomaterials using permittivity measurements can only be approximate when applying the Topp formula.\\

While the investigated models show great promise, particularly in the RF range, limitations in capturing low-frequency dispersion (e.g., in the CRIM model) underline the need for combining broadband dielectric methods with complementary low-frequency techniques, such as spectral induced polarization (SIP). Such combinations could improve parameter estimation for critical subsurface properties, including those in highly saline or organic-rich environments.\\

This study highlights promising correlations between dielectric measurements and clay mineral parameters. It demonstrates that ion exchange processes at clay surfaces significantly influence the dielectric spectrum in the RF range, enabling qualitative material property assessments. For instance, spatial heterogeneities in nutrient uptake capacity on agricultural fields could potentially be mapped. Additionally, in borehole applications, relationships between dielectric measurements and geomaterial parameters are already being utilized in industrial practices.\\

To deepen these insights, future research should include systematic variations in water content, porosity, and pore water salinity. These variations could help refine and evaluate existing models. However, experimental handling of highly swelling clays poses significant challenges and necessitates novel measurement and geotechnical techniques. Combining these methods with low-frequency approaches, such as spectral induced polarization (SIP) or complex conductivity (CC), could enhance accuracy and provide further valuable insights.\\

Expanding the experimental scope to include real soils with varying organic matter content and salinity would enhance the understanding of their dielectric behavior. As a next step, future research should aim to extend these analyses to real soils that contain soil organic matter. Particularly in surface soils, organic matter commonly contributes the largest share to the total cation exchange capacity (CEC), and its presence may significantly affect dielectric behavior. Addressing these influences would allow a more comprehensive and realistic modeling of natural soil systems. Organic matter, which strongly influences CEC, presents additional challenges due to its heterogeneity. Addressing these complexities could significantly improve the applicability of dielectric methods in natural and engineered geomaterials. This study also demonstrated the reliability of models for describing the dielectric properties of clays and clay-rich materials, which are critical for applications such as ground-penetrating radar (GPR). Furthermore, the limitations of the Topp formula in estimating soil moisture content in clay-rich soils were highlighted. By leveraging the proposed models, improvements in accuracy can be achieved, particularly in estimating water-filled porosity, a crucial parameter for hydrological and petrophysical applications.\\

Overall, the integration of experimental, modeling, and theoretical insights in this study paves the way for the development of robust, scalable, and non-invasive soil monitoring tools. These tools hold significant promise for improving agricultural productivity, optimizing water resource management, and supporting climate resilience strategies in the face of global environmental challenges.


\section{Open Research Section}
The broadband dielectric spectroscopy data and measured sample properties used in this study are openly available. The dataset includes scattering parameter files (.s2p) and measured mineralogical and petrophysical data. The data are available under \cite{Schmidt2025}.

\section{Acknowledgements}
Peter Weidler from the Institute of Functional Interfaces (IFG) at the Karlsruhe Institute of Technology (KIT) in Karlsruhe Germany for the measurement of the specific surface area. We would like to thank Yi-Yu Liu and Nadja Werling from the Competence Center for Material Moisture (CMM) KIT for measuring the CEC. We would like to thank Eleanor Bakker from CMM at the KIT for the characterisation of the mineral phases by X-ray diffraction.


\clearpage
\def\newblock{\ }%
\bibliographystyle{elsarticle-num}     
\bibliography{250508_broadband-dielectric-properties}

\end{document}